\documentclass[prd,aps,reprint,onecolumn,superscriptaddress,tightenlines,nofootinbib,eqsecnum,preprintnumbers,longbibliography,12pt]{revtex4-2}
\usepackage{bbm}
\usepackage{amsmath,amssymb}
\usepackage{physics}
\usepackage{dsfont}
\usepackage{graphicx}
\usepackage{hyperref}
\usepackage[dvipsnames]{xcolor}

\usepackage{bm}

\def\ps{\slash \! \! \! p}

\begin{document}

% Use the \preprint command to place your local institutional report
% number in the upper righthand corner of the title page in preprint mode.
% Multiple \preprint commands are allowed.
% Use the 'preprintnumbers' class option to override journal defaults
% to display numbers if necessary
%\preprint{}

%Title of paper
\title{Can CP be conserved in the two-Higgs-doublet model?}
\def\Carleton{Ottawa-Carleton Institute for Physics, Carleton University, Ottawa, ON K1S 5B6, Canada}
\def\TRIUMF{TRIUMF, 4004 Wesbrook Mall, Vancouver, BC V6T 2A3, Canada}

\author{Carlos Henrique de Lima}
\email{cdelima@triumf.ca}
\affiliation{\TRIUMF}
\affiliation{\Carleton}

\author{Heather E.\ Logan}
\email{logan@physics.carleton.ca}
\affiliation{\Carleton}

\date{\today}
\begin{abstract}
We study the conditions under which the CP violation in the quark mixing matrix can leak into the scalar potential of the real two-Higgs-doublet model (2HDM) via divergent radiative corrections, thereby spoiling the renormalizability of the model. We show that any contributing diagram must involve 12 Yukawa-coupling insertions and a factor of the $U(1)_{PQ}$-breaking scalar potential parameter $\lambda_5$, thereby requiring at least six loops; this also implies that the 2HDM with only softly-broken $U(1)_{PQ}$ is safe from divergent leaks of CP violation to all orders. We demonstrate that additional symmetries of the six-loop diagrams in the type I and II 2HDMs guarantee that all of the divergent CP-violating contributions cancel at this order. We also show that these symmetries are violated at seven loops and enumerate the classes of diagrams that can contribute to CP-violating divergences, providing evidence that the real 2HDM is theoretically inconsistent starting at the seven-loop level.
\end{abstract}

% insert suggested keywords - APS authors don't need to do this
%\keywords{}

%\maketitle must follow title, authors, abstract, and keywords
\maketitle

%%%%%%%%%%%%%%%%%%%%%%%%%%%%%%%%%%%%%%%%%%%
\section{Introduction}
%%%%%%%%%%%%%%%%%%%%%%%%%%%%%%%%%%%%%%%%%%%

Precision measurements of the 125~GeV Higgs boson discovered a decade ago~\cite{higgsdiscovery1,higgsdiscovery2} at the CERN Large Hadron Collider (LHC) indicate that it must be mainly composed of the real neutral component of a single scalar $SU(2)_L$ doublet~\cite{ATLAS:2022vkf,CMS:2022dwd}. Since there is no fundamental reason for the scalar sector to contain only a single Higgs doublet, it is essential to explore whether more complicated Higgs sectors can be consistent with existing data. Since the gauge representations of scalars that can contribute to electroweak symmetry breaking are heavily restricted via their modifications of the electroweak $\rho$ parameter~\cite{Langacker:1980js,ParticleDataGroup:2020ssz}, extended Higgs sectors that contain additional $SU(2)_L$ doublets are particularly well-motivated. The most straightforward of these extensions, consistent with current data, is the venerable two-Higgs-doublet model (2HDM)~\cite{Lee:1973iz}, first introduced in 1973 as a source of spontaneous CP violation.\footnote{CP denotes the combined symmetry transformation of charge conjugation and parity.}

The most popular versions of the 2HDM impose a softly-broken $Z_2$ symmetry on the Lagrangian in order to avoid experimentally-dangerous flavor-changing neutral Higgs couplings~\cite{Glashow:1976nt,Paschos:1976ay}, giving rise to the familiar type-I, -II, -X, and -Y 2HDMs (for recent reviews, see Refs.~\cite{Branco:2011iw,Ivanov:2017dad}). While these versions of the 2HDM can generally admit explicit CP violation in the scalar potential~\cite{Gunion:2005ja}, most studies impose the additional condition of CP conservation in the scalar potential, yielding the model known as the real 2HDM. This is increasingly convenient because experimental constraints on the electric dipole moment (EDM) of the electron~\cite{Roussy:2022cmp} are beginning to require fine-tuning of the CP-violating scalar potential parameter ${\rm Im}(\lambda_5)$ to be below $10^{-2}$, even for masses of the additional Higgs bosons at the TeV scale~\cite{Altmannshofer:2020shb}.

It was recently pointed out by Fontes et al.~\cite{Fontes:2021znm} that the real 2HDM is expected to be theoretically inconsistent under renormalization. CP violation in the Cabibbo-Kobayashi-Maskawa (CKM) quark-mixing matrix is experimentally well-established~\cite{ParticleDataGroup:2022pth}, and there is no apparent reason why this CP violation should not leak into the renormalized parameters of the scalar sector at high enough loop order. Since the CP violation in the CKM matrix ultimately arises from the dimensionless Yukawa couplings, this radiatively-induced CP breaking in the scalar potential should be hard; i.e., it should involve divergent contributions to dimension-four operators. Renormalizing the theory in the usual way would then require the presence of counterterms with imaginary parts, which are not present in the original scalar potential of the real 2HDM. This forces us to return to the complex 2HDM and face the (as-yet somewhat mild) fine-tuning of the CP violation in the scalar potential as well as its more complicated collider phenomenology~\cite{Keus:2015hva}, since there is no particular reason for this CP violation to be small. Such a leak would also provide the first known example in which a P-odd CP violation (from the CKM matrix) gives rise to a P-even CP violation (in the scalar potential)~\cite{Haber:2022gsn}.

The first step toward demonstrating that such a leak occurs was taken by Fontes et al.~\cite{Fontes:2021znm}, who performed an explicit computation of the coefficient of the leading $1/\epsilon^3$ divergence of the three-loop tadpole for the pseudoscalar $A^0$ in the type-II 2HDM. Three loops represent the lowest order at which the necessary four powers of the CKM matrix are present to generate the Jarlskog invariant~\cite{Jarlskog:1985cw}; a nonzero result would indicate the need for a counterterm for the vacuum expectation value (vev) of $A^0$, which is absent in the real 2HDM. Reference~\cite{Fontes:2021znm} found, unexpectedly, that after summing their result over all possible combinations of internal quark flavors, all contributions to the coefficient of this leading divergence exactly cancel. Our objective in this paper is to understand the source of this cancellation and to determine at what loop order a nonzero leak of CP violation should be expected to appear.

Our strategy in this paper is informed by the fact that identifying the divergences requiring counterterms at each loop order in the radiatively-corrected scalar potential is equivalent to identifying the contributions to the renormalization-group (RG) equations\footnote{The RG equations for the 2HDM up to three loops have been studied in Refs.~\cite{PhysRevD.56.5366,Fontes:2021iue,Oredsson:2019mni,Oredsson:2018yho,Herren:2017uxn,Bednyakov:2018cmx,Chowdhury:2015yja}.} that control the evolution of the corresponding parameters as a function of energy scale. Starting from a CP-conserving scalar potential at one scale, the presence of imaginary divergences arising from the CP violation in the CKM matrix would cause the imaginary parts of the scalar potential parameters to run away from zero, introducing CP violation at other energy scales.\footnote{From this point of view, one could in principle argue for a CP-conserving scalar potential at a particular matching energy scale due, e.g., to an accidental symmetry in the relevant sector of an underlying theory; if the running is numerically small, the CP violation thereby generated at the weak scale could then be argued to be small enough to be phenomenologically irrelevant.}
Because the RG equations for dimensionless Lagrangian parameters are not affected by spontaneous symmetry breaking or by the presence of dimensionful parameters such as masses, this approach allows us to work in the unbroken phase of the 2HDM and to ignore scalar masses when analyzing the contributing diagrams. This dramatically simplifies the analysis and highlights the critical roles played by natural flavor conservation (enforced by the softly-broken $Z_2$ symmetry) and other accidental symmetries in canceling the imaginary divergences at low loop orders.

In particular, the real 2HDM with natural flavor conservation contains only one source of CP violation, i.e., the Jarlskog invariant from quark mixing.\footnote{We ignore neutrino masses and the strong CP problem.}  In the unbroken phase, it becomes immediately apparent~\cite{Botella:1994cs} that 12 insertions of Yukawa matrices in a closed quark loop are necessary to construct the Jarlskog invariant. To generate a four-scalar operator from such a diagram requires already five loops, providing an immediate reason for the cancellation of the divergent part of the three-loop $A^0$ tadpole. It is also immediately apparent from this approach that the only four-scalar operators that can be generated from such a diagram at five loops are Hermitian, and to generate the non-Hermitian four-scalar operator corresponding to ${\rm Im}(\lambda_5)$ requires in addition an insertion of the tree-level $\lambda_5$ coupling, pushing the calculation to six loops. This is because, in the absence of $\lambda_5$, the model possesses an accidental global $U(1)$ Peccei-Quinn-type symmetry~\cite{Ferreira:2008zy} that is only softly broken by dimension-two terms, thereby guaranteeing that ${\rm Im}(\lambda_5)$ (and for that matter, ${\rm Re}(\lambda_5)$) is protected from divergent radiative corrections to all loop orders. 
As a bonus, we thus demonstrate that the CP-conserving 2HDM in which natural flavor conservation is enforced by the softly-broken $U(1)_{PQ}$ symmetry is guaranteed to be safe from divergent leaks of CP violation to all loop orders. Phenomenologically, this model differs from the usual softly-broken $Z_2$-symmetric 2HDM only in the strengths of some triple and quartic Higgs couplings, which affect, e.g., the charged Higgs contribution to the Higgs decay to two photons, and is thus fully phenomenologically viable up to now.

We then proceed to examine the six-loop diagrams in the type I and II 2HDMs.\footnote{Our analysis is not sensitive to the choices of couplings made in the lepton sector; thus, our results for the type I and type II models can be trivially extended to the type X (or Lepton-Specific) and type Y (or Flipped) models, respectively.}. We approach this problem by looking for symmetries or topological properties of the diagrams that could generate a cancellation of the imaginary part. Interestingly, this is precisely what happens at six loops in both the type I and type II models. In the type II 2HDM, the imaginary divergence cancels between pairs of diagrams due to an accidental generalized CP symmetry between the diagrams. In type I, the imaginary divergence also cancels, but in this case, it is because of the topology of the contributing diagrams: all diagrams involve a Hermitian five-loop subdiagram, and we can show that the dimensionless form factor associated with this subdiagram must be real.

Extending our analysis to the following order, it becomes immediately apparent that the arguments that provided a cancellation at six loops no longer hold at seven loops. In the type II model, the accidental generalized CP symmetry is broken by couplings that distinguish between the two Higgs doublets (such as $\lambda_1$ and $\lambda_2$) or between up- and down-type quarks (such as hypercharge or additional Yukawa insertions). Meanwhile, for type I, there appears a subset of diagrams that do not possess the subdiagram structure; these involve the quartic couplings $\lambda_{3}$ or $\lambda_{4}$ or hypercharge or $SU(2)_L$ gauge interactions. We thus expect imaginary divergent contributions to the scalar potential to appear at seven loops in both models. We emphasize that since we have not computed the seven-loop coefficient, additional as-yet-unidentified symmetries could lead to cancellations of the imaginary divergence even at the seven-loop level.

From our analysis, we expect the RG equation at seven loops for ${\rm Im}(\lambda_5)$ to have the form:
\begin{align}
 	\frac{d \, {\rm Im}(\lambda_{5})}{d \ln \mu} = 
 	\frac{ \lambda_5 {\rm Im}(\mathcal{J})}{(16 \pi^2)^7} \begin{cases} 
      \left[ a^{\lambda} (\lambda_1 - \lambda_2) 
			+ a^{g^{\prime}} g^{\prime 2} 
			+ a^y (y_t^2 - y_b^2 + \ldots) \right]   &  
       \, \text{(type II)} \\
      \\
      \left[ b^{\lambda_3} \lambda_3 + b^{\lambda_4} \lambda_4 + b^{g^{\prime}} g^{\prime 2} + b^g g^2 \right]  &    \, \text{(type I),} \\
   \end{cases}
\end{align}
where $\mu$ is the renormalization scale, ${\rm Im}(\mathcal{J})$ (defined in Sec.~\ref{sec:CPVbase}) is proportional to the Jarlskog invariant, $a^i$ and $b^i$ are numerical coefficients, $g^{\prime}$ and $g$ are the hypercharge and $SU(2)_L$ gauge couplings, $y_t$ and $y_b$ are the top- and bottom-quark Yukawa couplings, and the ellipses represent terms involving second- and first-generation quark Yukawa couplings. 

In Sec.~\ref{sec:CPVbase}, we estimate the size of ${\rm Im}(\mathcal{J})$ to be at most $\sim 8 \times 10^{-14}$ after taking into account possible $\tan\beta$ enhancement in the type II model. Taking, e.g., the quartic couplings to be of order one and the numerical factor $a^{\lambda}$ to be of order the number of contributing diagrams $\sim 3 \times 10^4$ (which we computed using QGRAF~\cite{QGRAF}), we find $d \, {\rm Im}(\lambda_5)/d \ln \mu \sim 10^{-24}$ at seven loops in the type II model. This is extraordinarily small, such that RG running from a CP-conserving scalar potential at the Planck scale down to the weak scale would generate a contribution to the electron EDM from ${\rm Im}(\lambda_5)$ more than ten orders of magnitude smaller than the estimated Standard Model (SM) contribution~\cite{Pospelov:2005pr} coming directly from the CKM matrix at four loops.

Nevertheless, if imaginary divergent contributions to scalar potential parameters arise at \emph{any} loop order, the real 2HDM is rendered inconsistent under renormalization. One must then question whether it makes sense to restrict the CP-violating scalar potential parameters to zero when the model structure does not preserve this decision.

The remainder of this paper is organized as follows. In Sec.~\ref{sec:2HDM}, we review the general 2HDM and the role of the global symmetries that give rise to natural flavor conservation and discuss their consequences for CP violation. In Sec.~\ref{sec:CPVbase}, we determine the combinations of Yukawa couplings that give rise to the Jarlskog invariant and construct the dimensionless combination $\mathcal{J}$ we use in our analysis. In Sec.~\ref{sec:6loop}, we analyze the six-loop contribution to the imaginary divergent part of $\lambda_5$ and show that it is zero due to symmetry relations among the contributing diagrams in the type I and II 2HDMs. In Sec.~\ref{sec:7loop}, we extend the analysis to seven loops and show that the symmetry arguments no longer hold at this order, and identify the parameter dependence of the potential imaginary divergences at this order. In Sec.~\ref{sec:m12sq}, we address the imaginary divergent contribution to the soft-$Z_2$-breaking parameter $m_{12}^2$ and show that it always appears at one loop order higher than the imaginary divergent contribution to $\lambda_5$. We conclude in Sec.~\ref{sec:conclusion}. In Appendix~\ref{sec:general2HDM}, we consider the general 2HDM without natural flavor conservation and illustrate the appearance of imaginary divergent contributions to $\lambda_5$ and $m_{12}^2$ already at the one-loop level. In Appendix~\ref{sec:tadpole}, we address the origin of the cancellation of the divergence in the three-loop $A^0$ tadpole in the broken phase of the 2HDM, thereby confirming the result found in Ref.~\cite{Fontes:2021znm}.

%%%%%%%%%%%%%%%%%%%%%%%%%%%%%%%%%%%%%%%%%%%
\section{Structure of the two-Higgs-doublet model}
\label{sec:2HDM}
%%%%%%%%%%%%%%%%%%%%%%%%%%%%%%%%%%%%%%%%%%%

%%%%%%%%%%%%%%%%%%%%%
\subsection{Most general 2HDM}
\label{sec:gen2HDM}
%%%%%%%%%%%%%%%%%%%%%%%%%%%%%%%%%%%%%%%%%%%

The most general 2HDM, constrained only by imposing $SU(2)_L \times U(1)_Y$ gauge invariance, is described by the scalar potential~\cite{Wu:1994ja,Branco:2011iw}:
\begin{eqnarray}
	V &=& m_{11}^2 \Phi_1^{\dagger} \Phi_1 + m_{22}^2 \Phi_2^{\dagger} \Phi_2
	- \left( m_{12}^2 \Phi_1^{\dagger} \Phi_2 + {\rm h.c.} \right) \nonumber \\
	&& + \frac{1}{2} \lambda_1 \left( \Phi_1^{\dagger} \Phi_1 \right)^2 
	+ \frac{1}{2} \lambda_2 \left( \Phi_2^{\dagger} \Phi_2 \right)^2 
	+ \lambda_3 \left( \Phi_1^{\dagger} \Phi_1 \right) \left( \Phi_2^{\dagger} \Phi_2 \right)
	+ \lambda_4 \left( \Phi_1^{\dagger} \Phi_2 \right) \left( \Phi_2^{\dagger} \Phi_1 \right) 
		\nonumber \\
	&& + \left[ \frac{1}{2} \lambda_5 \left( \Phi_1^{\dagger} \Phi_2 \right)^2
	+ \lambda_6 \left( \Phi_1^{\dagger} \Phi_1 \right) \left( \Phi_1^{\dagger} \Phi_2 \right)
	+ \lambda_7 \left( \Phi_2^{\dagger} \Phi_2 \right) \left( \Phi_1^{\dagger} \Phi_2 \right) 
	+ {\rm h.c.} \right],
	\label{eq:scalarpot}
\end{eqnarray}
where h.c.\ stands for Hermitian conjugate, and the doublets are given in terms of their component fields as $\Phi_i = (\phi_i^+, \phi_i^0)^T$. Here $m_{12}^2$, $\lambda_5$, $\lambda_6$, and $\lambda_7$ are in general complex parameters; after accounting for the various ways that phases can be absorbed into field redefinitions, their phases give rise to two independent physical CP-violating invariants~\cite{Lavoura:1994fv,Botella:1994cs} (see also Ref.~\cite{Gunion:2005ja}).

In this work, we focus on the quark Yukawa couplings since the CP violation from the CKM matrix originates there. Without imposing any further constraints, the two doublets $\Phi_1$ and $\Phi_2$ can each couple to both the down-type and the up-type right-handed quarks:
\begin{align}
\mathcal{L}_{Yuk} =&  - (Y_{d}^{(1)})_{ij} \overline{Q}_{Li}  \Phi_1 d_{Rj} 
	-  (Y_{u}^{(1)})_{ij} \overline{Q}_{Li} \widetilde{\Phi}_1 u_{Rj} + {\rm h.c.}  \nonumber \\
	& - (Y_{d}^{(2)})_{ij} \overline{Q}_{Li}  \Phi_2 d_{Rj} 
	-  (Y_{u}^{(2)})_{ij} \overline{Q}_{Li} \widetilde{\Phi}_2 u_{Rj} + {\rm h.c.}, \label{eq:genYuk}
\end{align}
where the flavor indices $i, j$ are summed over the three generations, and we define the conjugate doublet in terms of the antisymmetric tensor $\varepsilon$ as $\widetilde \Phi_i \equiv \varepsilon \Phi_{i}^* = (\phi_i^{0*}, -\phi_i^{+*})^T$.

After electroweak symmetry breaking, these Yukawa couplings lead generically to unacceptably large flavor-changing neutral Higgs interactions and new sources of CP violation beyond that contained in the CKM matrix.  
While the parameters $m_{12}^2$, $\lambda_5$, $\lambda_6$, and $\lambda_7$ can be chosen to be real at tree level, this model admits leaks of CP violation from the Yukawa couplings to the scalar potential already at one loop. We demonstrate this explicitly in Appendix~\ref{sec:general2HDM}. This situation is similar to the toy model of CP leaks presented in Ref.~\cite{Fontes:2021znm}, in which both Higgs doublets are coupled to the same additional scalar fields via CP-violating couplings, and the divergent leak occurs at one loop. It is also consistent with the results of Ref.~\cite{Cao:2022rgh}, which found via a geometric argument that the additional phases in these extra Yukawa couplings generically introduce CP violation into the one-loop effective potential of the 2HDM.
We will see in what follows that the situation is dramatically different after imposing natural flavor conservation, which protects the scalar sector from leaks of CP violation at low loop orders.

%%%%%%%%%%%%%%%%%%%%%%%%%%%%%%%%%%%%%%%%%%%
\subsection{Natural flavor conservation and global symmetry}
%%%%%%%%%%%%%%%%%%%%%%%%%%%%%%%%%%%%%%%%%%%

To avoid the phenomenological problems associated with flavor-changing neutral Higgs interactions, one usually implements {\it natural flavor conservation}~\cite{Glashow:1976nt,Paschos:1976ay} by imposing an additional (global) symmetry on the theory to force the right-handed fermions of each electric charge to couple to only a single Higgs doublet. The most familiar implementation involves imposing a $Z_2$ symmetry, under which
\begin{equation}
	\Phi_1 \to -\Phi_1, \qquad \Phi_2 \to \Phi_2,
\end{equation}
and the quarks transform according to\footnote{The various possible assignments of the $Z_2$ transformation properties of the right-handed charged leptons distinguish between types I and X and between types II and Y; our analysis in what follows is independent of these choices.}
\begin{eqnarray}
	&&u_R \to u_R, \qquad d_R \to d_R,  \qquad \ \ {\rm (type~I)} \nonumber \\
	&&u_R \to u_R,  \qquad d_R \to -d_R,  \qquad {\rm (type~II)}
\end{eqnarray}
with $Q_L$ invariant.

The $Z_2$ symmetry forces the Yukawa couplings to take the following form in the type II model (the type I model is obtained by replacing $\Phi_1$ with $\Phi_2$ in the first term):
\begin{equation}
	\mathcal{L}_{Yuk} = - (Y_d)_{ij} \overline{Q}_{Li} \Phi_1 d_{Rj} 
		- (Y_u)_{ij} \overline{Q}_{Li} \widetilde \Phi_2 u_{Rj} + {\rm h.c.},
	\label{eq:LYuk}
\end{equation}
and forces $m_{12}^2 = \lambda_6 = \lambda_7 = 0$ in the scalar potential.  This leaves $\lambda_5$ as the only complex parameter in the scalar potential; however, its phase can always be rotated away by a suitable field redefinition and is not physical. Thus, this model cannot have an explicit CP violation in the scalar potential, and this fact is enforced to all orders by the exact $Z_2$ symmetry: in the $Z_2$-symmetric 2HDM without soft breaking, there can be no leak of CP violation from the Yukawa sector.

However, this model lacks a decoupling limit~\cite{Haber:1994mt},\footnote{In fact, the exact $Z_2$ symmetry limit of the type II, X, and Y 2HDMs is now excluded by a global Bayesian fit to LHC data in each model~\cite{Chowdhury:2017aav}, and the type I 2HDM is heavily constrained. This is a general trend for non-decoupling extensions of the Higgs sector as we increase the experimental precision of Higgs couplings and searches for new scalars~\cite{deLima:2022yvn}.} so the $Z_2$ symmetry is usually allowed to be softly broken by a nonzero $m_{12}^2$. With the soft breaking, radiative corrections can give rise to \emph{finite} wrong-Higgs Yukawa couplings at the two-loop level and finite four-Higgs operators corresponding to the $\lambda_6$ and $\lambda_7$ couplings; however, they cannot give rise to divergent contributions to these operators. The model thus remains renormalizable since all divergences can be canceled by the counterterms present in the original Lagrangian.

Once $m_{12}^2$ is allowed in the theory, the scalar potential generically contains CP violation since, in general, the phases of $m_{12}^2$ and $\lambda_5$ cannot be simultaneously rotated away. The scalar potential then contains one physically meaningful complex phase, given by ${\rm Phase}[(m_{12}^{2*})^2 \lambda_5]$. This is the model known as the complex 2HDM or C2HDM~\cite{Ginzburg:2002wt} (for a recent phenomenological review, see also Ref.~\cite{Fontes:2017zfn}); its one-loop electroweak renormalization was developed in Ref.~\cite{Fontes:2021iue}.  

The imposition of CP conservation in the scalar potential constitutes choosing $m_{12}^2$ and $\lambda_5$ to be real in the same basis; i.e., choosing ${\rm Im}[(m_{12}^{2*})^2 \lambda_5] = 0$. This model is the CP-conserving or real 2HDM and is the version of the 2HDM we consider in the remainder of this paper.

%%%%%%%%%%%%%%%%%%%%%%%%%%%%%%%%%%%%%%%%%%%
\subsection{An alternative global symmetry}
%%%%%%%%%%%%%%%%%%%%%%%%%%%%%%%%%%%%%%%%%%%

The Yukawa couplings in Eq.~(\ref{eq:LYuk}) are also invariant under a global $U(1)$ symmetry~\cite{Wilczek:1977pj,Weinberg:1977ma,Ferreira:2010ir},
\begin{equation}
	\Phi_1 \to e^{-i \theta} \Phi_1, \qquad \Phi_2 \to e^{i \theta} \Phi_2,
	\label{eq:U1PQ}
\end{equation}
with the quarks transforming according to
\begin{eqnarray}
	&&u_R \to e^{i \theta} u_R, \qquad d_R \to e^{-i \theta} d_R \qquad {\rm (type~I)} \nonumber \\
	&&u_R \to e^{i \theta} u_R, \qquad d_R \to e^{i \theta} d_R \qquad \ \ {\rm (type~II)},
\end{eqnarray}
with $Q_L$ invariant.  In the type II case, this global $U(1)$ is equivalent to the Peccei-Quinn symmetry~\cite{Peccei:1977hh,Wilczek:1977pj,Weinberg:1977ma}, so we will refer to it (in all cases) as $U(1)_{PQ}$.

This $U(1)_{PQ}$ symmetry can be imposed instead of $Z_2$ to implement natural flavor conservation. Extending it to the Higgs scalar potential forces $m_{12}^2 = \lambda_5 = \lambda_6 = \lambda_7 = 0$~\cite{Ferreira:2008zy} -- i.e., all of the scalar potential parameters that could be complex are set to zero. The scalar potential is then strictly CP conserving. 

If electroweak symmetry breaking leads both doublets to acquire nonzero vevs, the $U(1)_{PQ}$ is spontaneously broken and gives rise to a physical massless Goldstone boson; i.e., the pseudoscalar $A^0$ is massless.\footnote{In the type II model, $A^0$ is the original QCD axion~\cite{Wilczek:1977pj,Weinberg:1977ma} and acquires a small mass from QCD instanton effects of order $m_{\pi} f_{\pi}/v \sim {\rm MeV}$, which is thoroughly excluded by meson decays.  A version of this model with gauged $U(1)_{PQ}$ was studied in Ref.~\cite{Davoudiasl:2012ag}; in that case, the would-be Goldstone boson is eaten by the $U(1)_{PQ}$ gauge boson and the model is viable.}  This is phenomenologically unviable. This problem can be solved by allowing the $U(1)_{PQ}$ to be softly broken by $m_{12}^2$, thereby giving $A^0$ a mass and allowing a decoupling limit as in the $Z_2$ case. This softly broken version of the model is phenomenologically viable, differing from the usual softly-broken $Z_2$-symmetric 2HDM only through restrictions on the strengths of certain triple- and quartic-scalar couplings associated with the restriction $\lambda_5 = 0$. These couplings affect, e.g., the charged Higgs loop contribution to $h^0 \to \gamma\gamma$ and some Higgs-to-Higgs decays of the heavier scalars.

Even with the soft breaking term, the $U(1)_{PQ}$ symmetry still forces $\lambda_5 = 0$. With $m_{12}^2$ being the only complex parameter allowed in the scalar potential, field redefinitions can rotate away its phase, and the scalar potential is then CP conserving. The model is renormalizable, so the counterterms of the existing Lagrangian parameters must cancel all divergent radiative corrections.   In this model, therefore, the $U(1)_{PQ}$ symmetry guarantees no divergent leaks of CP violation from the CKM matrix into the scalar sector.

%%%%%%%%%%%%%%%%%%%%%%%%%%%%%%%%%%%%%%%%%%%
\subsection{Consequences for renormalization group equations}
\label{sec:symcons}
%%%%%%%%%%%%%%%%%%%%%%%%%%%%%%%%%%%%%%%%%%%

The symmetries we discussed in the preceding subsections are reflected in the form that the RG equations must take. We focus here on the 2HDM with softly broken $Z_2$ symmetry. Aside from the Lagrangian parameters $\lambda_5$ and $m_{12}^2$, the entire Lagrangian (and particularly, the Yukawa couplings) preserves a $U(1)_{PQ}$ symmetry. Among the dimension-zero couplings, only $\lambda_5$ breaks the $U(1)_{PQ}$ down to $Z_2$.  Therefore, any divergent radiative correction to $\lambda_5$ must know about the presence of nonzero $\lambda_5$ in the original scalar potential. The breaking of $U(1)_{PQ}$ cannot be conveyed to the dimensionless coupling $\lambda_5$ solely through factors of $m_{12}^2$ since the latter is a soft breaking coupling~\cite{Symanzik:1970zz}. Similarly, $m_{12}^2$ is the only Lagrangian parameter that breaks the $Z_2$ symmetry; therefore, any divergent radiative correction to $m_{12}^2$ must know about the presence of nonzero $m_{12}^2$ in the original scalar potential.\footnote{This also ensures that the radiative corrections to $m_{12}^2$ are at most logarithmically divergent, not quadratically divergent.}

One further restriction on the form of the RG equations for $\lambda_5$ and $m_{12}^2$ can be obtained from the phase redefinitions of the fields that redistribute the complex phases of these two parameters. That phase redefinition can be chosen without loss of generality to be the $U(1)_{PQ}$ symmetry transformation of Eq.~(\ref{eq:U1PQ}). Under this transformation, $\lambda_5$ and $m_{12}^2$ transform according to 
\begin{equation}
	\lambda_5 \to e^{4 i \theta} \lambda_5, \qquad m_{12}^2 \to e^{2 i \theta} m_{12}^2,
\end{equation}
so that the combination $(m_{12}^{2*})^2 \lambda_5$ is invariant.  The RG equations for $\lambda_5$ and $m_{12}^2$ must transform under this symmetry like the original parameters. Together with the soft-breaking arguments above, this implies that all terms in the RG equation for $\lambda_5$ must be proportional to $\lambda_5$ (and not $\lambda_5^*$). Additional $|\lambda_5|^2$ powers are allowed at higher orders. Similarly, all terms in the RG equation for $m_{12}^2$ must be proportional to $m_{12}^2$ or $\lambda_5 m_{12}^{2*}$.  Again, additional powers of dimensionless combinations of couplings are allowed at higher orders as long as those combinations are invariant under $U(1)_{PQ}$.

These symmetry considerations have significant consequences for the loop order at which a divergent leak of CP violation can enter the scalar potential. In particular, an imaginary divergent contribution to $\lambda_5$ \emph{cannot} arise from a diagram involving only the Yukawa couplings because the Yukawa couplings by themselves preserve the $U(1)_{PQ}$ symmetry. Instead, any contributing diagram must include at least one quartic scalar vertex to provide a factor of $\lambda_5$. This requirement becomes apparent immediately when one begins trying to draw diagrams in the unbroken phase that produce the $\mathcal{O}_5 \equiv \Phi_1^{\dagger} \Phi_2 \Phi_1^{\dagger} \Phi_2$ operator.

%%%%%%%%%%%%%%%%%%%%%%%%%%%%%%%%%%%%%%%%%%%
\section{CP violation from the Yukawa matrices}
\label{sec:CPVbase}
%%%%%%%%%%%%%%%%%%%%%%%%%%%%%%%%%%%%%%%%%%%

All CP-violating observables arising from the CKM matrix $V$ can be parameterized in terms of a single quantity known as the Jarlskog invariant~\cite{Jarlskog:1985ht,Jarlskog:1985cw}, given in terms of the elements of $V$ by
\begin{equation}
	J = \left| {\rm Im} ( V_{\alpha i} V_{\beta j} V^*_{\alpha j} V^*_{\beta i} ) \right| 
	\qquad (\alpha \neq \beta, i \neq j).
\end{equation}

Expressing this invariant in terms of the Yukawa matrices will be most convenient for our purposes. Following Botella and Silva~\cite{Botella:1994cs} (see also Ref.~\cite{Silva:2004gz}), in the SM, we can define Hermitian combinations of up- and down-type quark Yukawa matrices according to
\begin{eqnarray}
	H_u &=& \frac{v^2}{2} Y_u Y_{u}^{\dagger} = U_{u_L} M_U^2 U_{u_L}^\dagger, \nonumber \\
	H_d &=& \frac{v^2}{2} Y_d Y_{d}^{\dagger} = U_{d_L} M_D^2 U_{d_L}^\dagger,
\end{eqnarray}
where $v \simeq 246~{\rm GeV}$ is the SM Higgs vacuum expectation value (vev), $Y_{u,d}$ are the up-and down-type quark Yukawa matrices, $M_{U,D}$ are the diagonal up- and down-type quark mass matrices in the mass basis, and the matrices $U$ are the unitary rotation matrices applied to the left-handed quarks\footnote{The corresponding rotation matrices applied to the right-handed quarks cancel in the combination $Y_q Y_q^{\dagger}$.} in the diagonalization of the mass matrices. The minimal combination of these Yukawa matrices that yields the Jarlskog invariant is then given by~\cite{Botella:1994cs,Roldan}

\begin{eqnarray}
	\overline J &=& {\rm Im} \left\{ {\rm Tr} \left( H_u H_d H_u^2 H_d^2 \right) \right\} \nonumber \\
	&=& {\rm Im} \left\{ {\rm Tr} \left( V^{\dagger} M_U^2 V M_D^2 V^{\dagger} M_U^4 V M_D^4 \right) \right\} \nonumber \\
	&=& T(M_U^2) \, B(M_D^2) \, J,
	\label{eq:Jbar}
\end{eqnarray}
where the functions of masses $T(M_U^2)$ and $B(M_D^2)$~\cite{Jarlskog:1985cw}, 
\begin{eqnarray}
	T(M_U^2) &=& m_t^4 (m_c^2 - m_u^2) + m_c^4 (m_u^2 - m_t^2) + m_u^4 (m_t^2 - m_c^2), \nonumber \\
	B(M_D^2) &=& m_b^4 (m_s^2 - m_d^2) + m_s^4 (m_d^2 - m_b^2) + m_d^4 (m_b^2 - m_s^2),
	\label{eq:TB}
\end{eqnarray}
are antisymmetric under the interchange of any pair of quark masses.

The trace in Eq.~(\ref{eq:Jbar}) contains the minimal combination of Yukawa matrices that can give rise to an imaginary part (and hence to a CP-violating result). This can be demonstrated by considering the complex conjugate of the traces of various combinations of $H_u$ and $H_d$ and using the cyclic property of the trace and the fact that $H_{u,d}$ are Hermitian by construction to attempt to reduce it to the original trace. It is easy to show in this way that ${\rm Tr}(H_u H_d H_u H_d)$ is purely real and that the combination in Eq.~(\ref{eq:Jbar}) is the lowest-order combination that can give an imaginary part. The combinations ${\rm Tr}(H_u^3 H_d^3)$ and ${\rm Tr}(H_u H_d H_u H_d H_u H_d)$, which are the same order in Yukawa insertions as the combination in Eq.~(\ref{eq:Jbar}), are both purely real by the same method. More generally, the combination ${\rm Tr}(H_u^{n_1} H_d^{m_1} H_u^{n_2} H_d^{m_2})$ can also be shown to yield an imaginary part proportional to the Jarlskog invariant so long as $n_1 \neq n_2$ and $m_1 \neq m_2$.

The form in Eq.~(\ref{eq:Jbar}) points the way to defining a useful CP-violating quantity when considering scalars attached to a closed quark loop in the unbroken phase of a multi-Higgs-doublet model with natural flavor conservation. In the unbroken phase, all of the CKM-related CP breaking is contained in the Yukawa matrices, and we can ignore the gauge sector entirely. We first remove the SM vev dependence from the definitions of Eq.~(\ref{eq:Jbar}) such that the notation also makes sense in the unbroken phase:
\begin{equation}
	\widehat{H}_u =  Y_u Y_{u}^{\dagger}, \qquad \qquad
	\widehat{H}_d =  Y_d Y_{d}^{\dagger}.
\end{equation}
We then define the following dimensionless combination of Yukawa couplings:
\begin{align} \label{def:complex}
\mathcal{J}= \Tr \left( \widehat{H}_u  \widehat{H}_d  \widehat{H}_u^2  \widehat{H}_d^2 \right) \, .
\end{align}
In the unbroken phase, traces of this form arise naturally from the sums over quark flavors in individual Feynman diagrams containing scalars attached to a closed quark loop.
The imaginary part of $\mathcal{J}$ is related to the usual SM Jarlskog invariant after electroweak symmetry breaking. In the type II 2HDM, we can write,
\begin{equation}
	{\rm Im}(\mathcal{J}) = \frac{1}{\sin^6 \beta \cos^6 \beta} \left( \frac{2}{v^2} \right)^6 
		T(M_U^2) B(M_D^2) J,
		\label{eq:ImJ}
\end{equation}
where $\tan\beta \equiv v_2/v_1$ is the usual ratio of doublet vevs. In the type I model, $\sin^6 \beta \cos^6 \beta$ is replaced by $\sin^{12} \beta$.  ${\rm Im}(\mathcal{J})$ is numerically quite small; evaluating Eq.~(\ref{eq:ImJ}) in terms of the SM parameters~\cite{ParticleDataGroup:2022pth} and estimating the QCD corrections to the Yukawa couplings by using the running quark masses evaluated at the SM Higgs mass scale in the modified minimal subtraction ($\overline{\rm MS}$) scheme~\cite{Huang:2020hdv} yields ${\rm Im}(\mathcal{J}) \simeq 2 \times 10^{-24}/\sin^6\beta \cos^6\beta$. Taking $\tan\beta \sim 60$ in the type II model to enhance the down-type quark Yukawa couplings as much as possible gives a maximum value of ${\rm Im}(\mathcal{J}) \sim 8 \times 10^{-14}$.

\begin{figure}[h!]
 \resizebox{0.28\linewidth}{!}{ \includegraphics{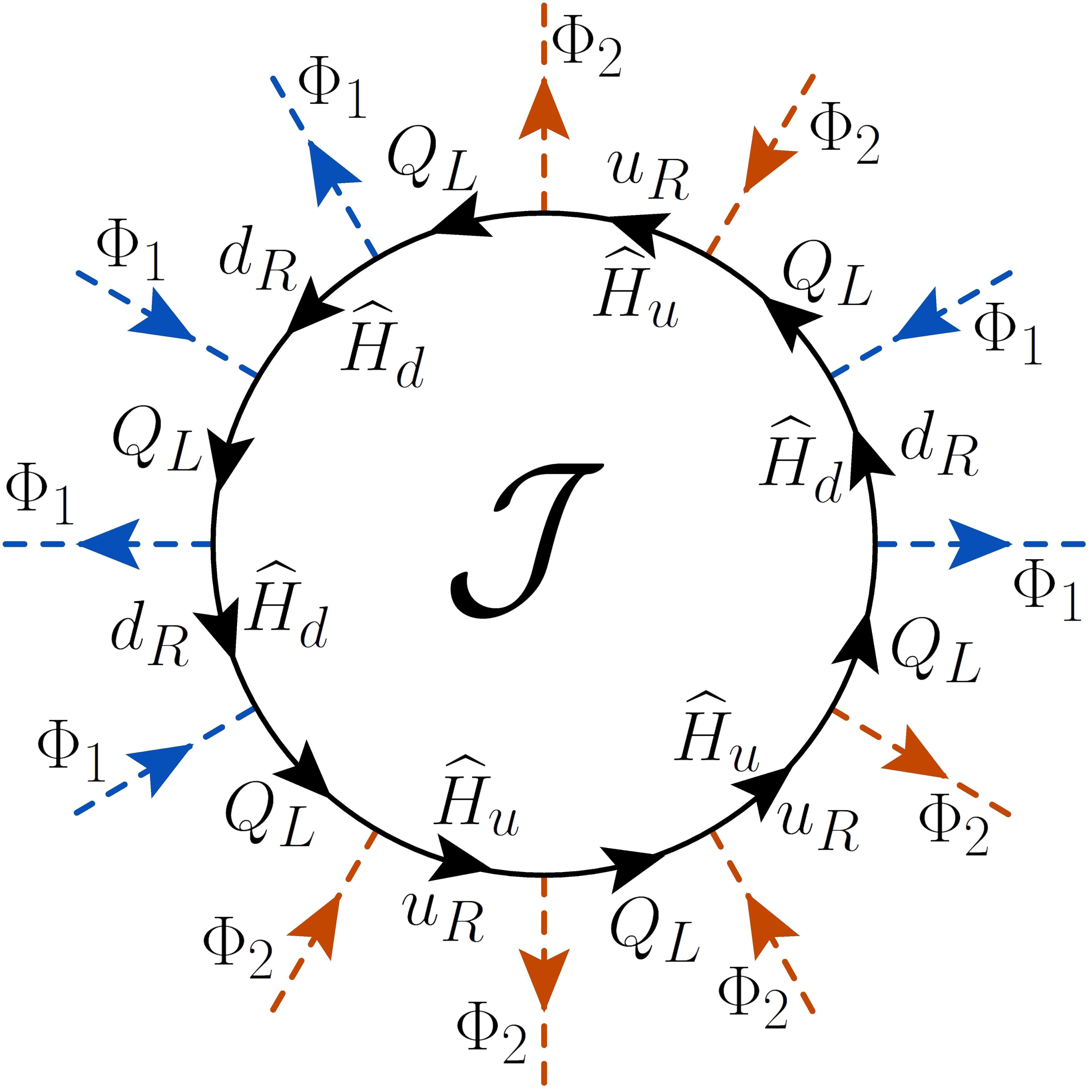}}
\caption{Primitive diagram proportional to $\mathcal{J}$ in the type II 2HDM.  Type I is obtained by replacing each $\Phi_1$ with $\Phi_2$. The primitive diagram proportional to $\mathcal{J}^*$ can be obtained by reversing the fermion flow or interchanging $u_{R} \leftrightarrow d_{R}$. }
\label{fig:bigcircle}
\end{figure}

In particular, the minimal combination of Yukawa matrices that can give rise to a CP-violating result involves six powers each of up-type and down-type Yukawa matrices, i.e., twelve Yukawa matrix insertions in the closed quark loop. This structure is illustrated in Fig.~\ref{fig:bigcircle} for the type II 2HDM (type I can be recovered by replacing each $\Phi_1$ with $\Phi_2$). Notice that since the Yukawa interactions preserve $U(1)_{PQ}$, any diagram generated only from these interactions is itself guaranteed to preserve $U(1)_{PQ}$; thus, the diagram in Fig.~\ref{fig:bigcircle} involves an equal number of ingoing and outgoing $\Phi_1$ fields, and similarly for $\Phi_2$. One can then immediately see that pairing up eight of the scalar legs of this diagram to produce a four-scalar operator will never produce $\mathcal{O}_5 \equiv \Phi_1^{\dagger} \Phi_2 \Phi_1^{\dagger} \Phi_2$ in either the type I or type II model; instead, an additional $\lambda_5$ vertex must be added, as we showed using symmetry arguments in Sec.~\ref{sec:symcons}. This means that the minimum number of loops required to have a chance to produce an imaginary divergent contribution to $\lambda_5$ from the CKM phase is six.

The primitive diagram in Fig.~\ref{fig:bigcircle}, together with the complex conjugated diagram proportional to $\mathcal{J}^*$, is the basis for all Feynman diagrams that can potentially give rise to imaginary divergent contributions to scalar potential parameters. Starting at six loops and considering the contributions to $\mathcal{O}_{5}$, we can then look for relations between the diagrams involving $\mathcal{J}$ and $\mathcal{J}^*$ and see if their contributions to the imaginary divergent part cancel. This allows us to identify cancellations without computing the divergent pieces of the actual diagrams. The key trick to finding such pairs of diagrams is to look for transformations that take $\mathcal{J}$ to $\mathcal{J}^*$, which are as follows:
\begin{enumerate}
\item Reversing the flow of fermion number. This reverses the flow of each scalar line and is equivalent to the charge conjugation of the entire diagram.
\item Interchanging $u_R$ and $d_R$ in the fermion line.  In the type I model, this interchanges $\Phi_2 \leftrightarrow \widetilde \Phi_2$, equivalent to charge conjugation on the scalars. In the type II model, this interchanges $\Phi_1 \leftrightarrow \widetilde \Phi_2$, which is equivalent to a generalized CP transformation\footnote{For a review of generalized CP transformations in the context of 2HDMs, see Sec.~5.5 of Ref.~\cite{Branco:2011iw}.} on the scalars.
\end{enumerate}
In the following sections, we use these transformations to identify pairs of diagrams whose imaginary divergent parts cancel.

Finally, one can show that the vanishing of the divergent piece of the three-loop $A^0$ tadpole computed in the broken phase in Ref.~\cite{Fontes:2021znm} is due to there not being enough powers of quark masses in the numerator to give rise to the antisymmetric combinations that must multiply the Jarlskog invariant. While additional powers of quark masses can be pulled from expansions of the quark propagators, this always gives rise to symmetric pairs of terms between which the imaginary part of the product of CKM matrix elements cancels. We provide some details of this cancellation in Appendix~\ref{sec:tadpole}.

%%%%%%%%%%%%%%%%%%%%%%%%%%%%%%%%%%%%%%%%%%%
\section{$\lambda_5$ at six loops}
\label{sec:6loop}
%%%%%%%%%%%%%%%%%%%%%%%%%%%%%%%%%%%%%%%%%%%

Having determined that any imaginary divergent contribution to $\lambda_5$ must be proportional to $\lambda_5 {\rm Im}(\mathcal{J})$, we now proceed to examine the six-loop diagrams that contain this combination of couplings. We work in the unbroken phase of the 2HDM, in which $Q_L$, $u_R$, and $d_R$ are separate massless degrees of freedom. Furthermore, since the RG equations of dimensionless couplings (including $\lambda_5$) cannot depend on dimensionful parameters~\cite{Weinberg:1973xwm}, we can ignore the mass-squared parameters $m_{11}^2$, $m_{22}^2$, and $m_{12}^2$ in this part of the analysis\footnote{We address the imaginary divergent contributions to $m_{12}^2$ in Sec.~(\ref{sec:m12sq}).}, and treat the doublets $\Phi_1$ and $\Phi_2$ as massless scalar degrees of freedom. We consider only the ultraviolet divergences in what follows. Infrared divergences can be regulated by adding a common auxiliary mass to propagators as needed~\cite{Chetyrkin:1997fm}; since this procedure does not affect the local ultraviolet divergences, and any nonlocal ultraviolet divergences (due to divergent subdiagrams) are guaranteed to be canceled by lower-order counterterms, this does not affect our results.

We examine the type II and type I 2HDMs separately. In both cases, we demonstrate that all imaginary divergent contributions to $\lambda_5$ at six-loop order cancel in the sum over contributing diagrams.

%%%%%%%%%%%%%%%%%%%%%%%%%%%%%%%%%%%%%%%%%%%
\subsection{Type II 2HDM}
\label{sec:6loopTypeII}
%%%%%%%%%%%%%%%%%%%%%%%%%%%%%%%%%%%%%%%%%%%

We begin with the type II 2HDM because the cancellation of the imaginary divergent contributions to $\lambda_5$ at six loops is more transparent in this model. The proof is as follows.

Consider an individual six-loop diagram generating the operator $\mathcal{O}_5 \equiv \Phi_1^{\dagger} \Phi_2 \Phi_1^{\dagger} \Phi_2$ that is proportional to $\mathcal{J} = {\rm Tr}(\widehat H_u \widehat H_d \widehat H_u^2 \widehat H_d^2)$.  This diagram can be written as
\begin{equation}
	\mathcal{M}_i^{(6)} \mathcal{O}_i = 
	\left\{ {\rm Tr}(\widehat H_u \widehat H_d \widehat H_u^2 \widehat H_d^2) 
	\mathcal{N}^i \right\}
	\Phi_1^{\dagger}(p_1) \Phi_2(p_2) \Phi_1^{\dagger}(p_3) \Phi_2(p_4),
\end{equation}
where the $SU(2)_L$ index structure of the doublets has been captured in the contractions of the four external scalar fields. The factor $\mathcal{N}^i$ contains the rest of the loop integral and is a function of the incoming four-momenta $p_{1 \cdots 4}$ and the scalar masses.

For each such diagram $i$, there exists a second diagram $i^{\prime}$, with identical topology and momentum structure but in which each $u_R$ is replaced with $d_R$ and each $d_R$ is replaced with $u_R$. Interchanging $u_R \leftrightarrow d_R$ in this way requires replacing $\Phi_1 \leftrightarrow \widetilde \Phi_2$ and $Y_u \leftrightarrow Y_d$.  All other aspects of the diagram remain the same, including the $SU(2)_L$ index structure, which we can show as follows.

\begin{figure}[h!]
 \resizebox{0.26\linewidth}{!}{ \includegraphics{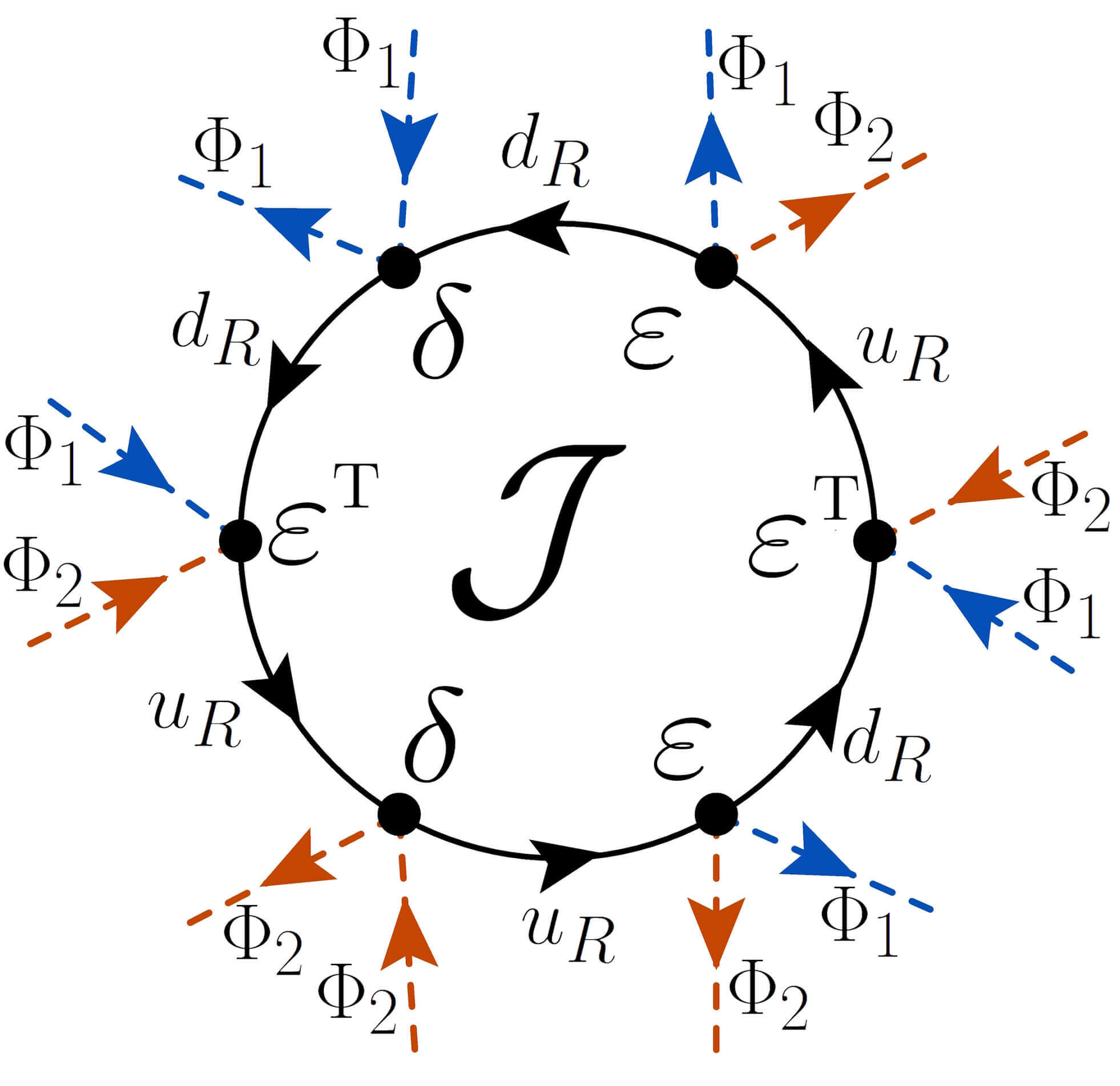}} \, \,  \, \, \, \, \resizebox{0.26\linewidth}{!}{ \includegraphics{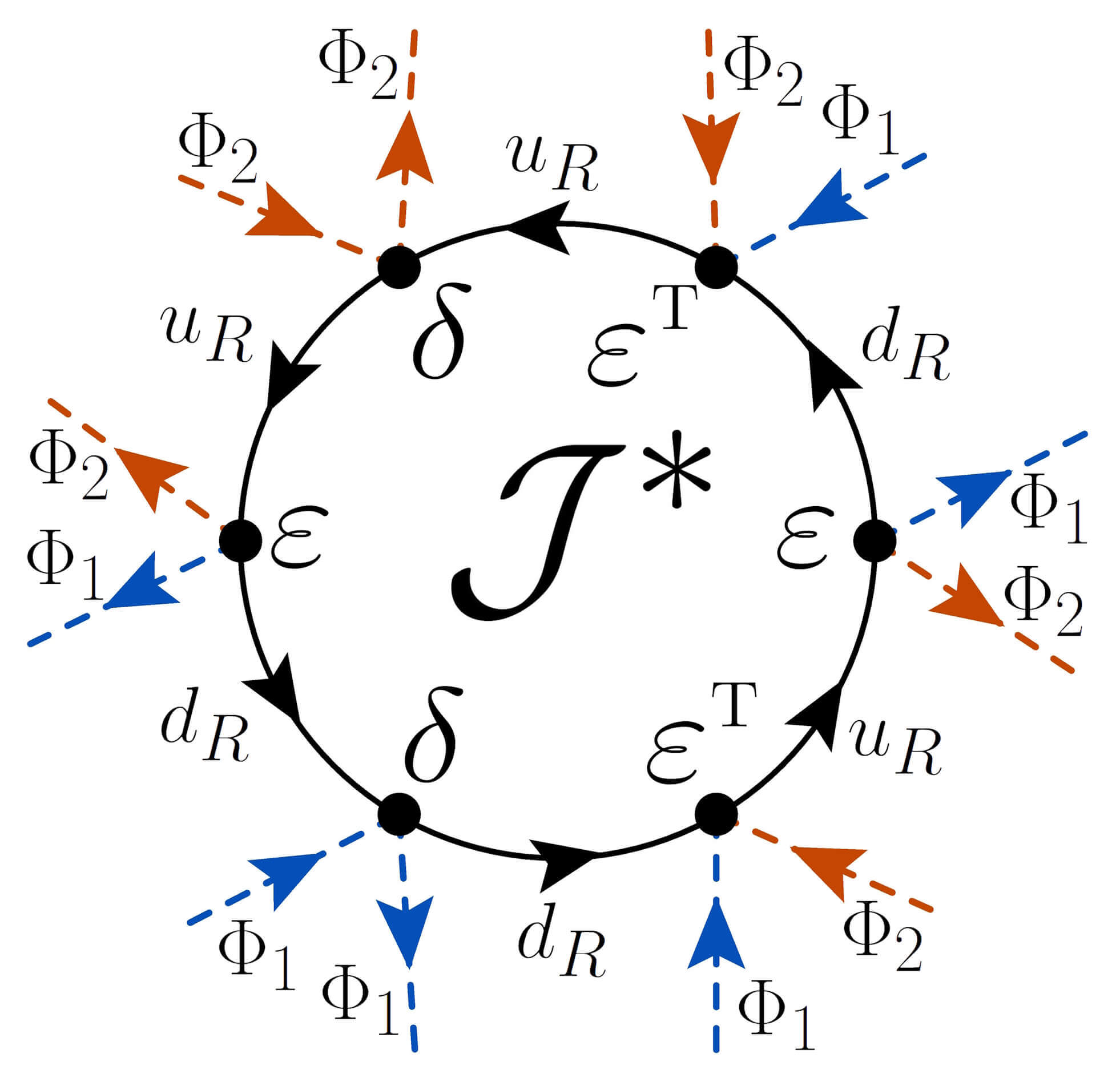}}
\caption{$SU(2)_{L}$ structure for the primitive $\mathcal{J}$ (left) and $\mathcal{J}^{*}$ (right) diagrams.}
\label{fig:eps}
\end{figure}

Following the fermion line backward as usual, we will denote the $SU(2)_L$ index of the $n$th attached Higgs doublet with $j_n$ and that of the left-handed quark doublet $Q_L$ connected to the same Yukawa vertex with $i_n$. Each Yukawa vertex involving $d_R$ and each $Q_L$ propagator gives rise to a Kronecker delta $\delta$, while each Yukawa vertex involving $u_R$ gives rise to an antisymmetric tensor $\varepsilon$. The $SU(2)_L$ structure of the Higgs doublets in diagram $i$ is then given by
\begin{eqnarray}
	&&(\varepsilon_{i_2j_2}\delta_{i_2i_3}\delta_{i_3j_3}) 
	(\delta_{i_4j_4}\delta_{i_4i_5}\varepsilon_{i_5j_5}) 
	(\varepsilon_{i_6j_6}\delta_{i_6i_7}\varepsilon_{i_7j_7})
	(\varepsilon_{i_8j_8}\delta_{i_8i_9}\delta_{i_9j_9}) 
	(\delta_{i_{10}j_{10}}\delta_{i_{10}i_{11}}\delta_{i_{11}j_{11}}) 
	(\delta_{i_{12}j_{12}} \delta_{i_{12}i_1}\varepsilon_{i_1j_1})  \nonumber \\
	&& \qquad = (-\varepsilon_{j_2j_3})(\varepsilon_{j_4j_5})(\delta_{j_6j_7})
	(-\varepsilon_{j_8j_9})(\delta_{j_{10}j_{11}})(\varepsilon_{j_{12}j_1}).
\end{eqnarray}
Meanwhile, the $SU(2)_L$ structure of the Higgs doublets in diagram $i^{\prime}$ is given by
\begin{eqnarray}
	&& (\delta_{i_2j_2}\delta_{i_2i_3}\varepsilon_{i_3j_3})
	(\varepsilon_{i_4j_4}\delta_{i_4i_5}\delta_{i_5j_5})
	(\delta_{i_6j_6}\delta_{i_6i_7}\delta_{i_7j_7})
	(\delta_{i_8j_8}\delta_{i_8i_9}\varepsilon_{i_9j_9})
	(\varepsilon_{i_{10}j_{10}}\delta_{i_{10}i_{11}}\varepsilon_{i_{11}j_{11}})
	(\varepsilon_{i_{12}j_{12}}\delta_{i_{12}i_1}\delta_{i_1j_1}) \nonumber \\
	&& \qquad = (\varepsilon_{j_2j_3}) (-\varepsilon_{j_4j_5}) (\delta_{j_6j_7}) 
	(\varepsilon_{j_8j_9}) (\delta_{j_{10}j_{11}}) (-\varepsilon_{j_{12}j_1}),
\end{eqnarray}
which is identical to the structure in diagram $i$. This feature occurs because going from $\mathcal{J}$ to $\mathcal{J}^*$ replaces all $\varepsilon \leftrightarrow \varepsilon^{T}$ and the structure contains an even number of both $\varepsilon$ and $\varepsilon^T$. This is illustrated diagrammatically in Fig.~\ref{fig:eps} after summing over the repeated $SU(2)_L$ indices of the $Q_L$.

We thus demonstrate that $\mathcal{N}^{i^\prime} = \mathcal{N}^i$. Diagram $i^{\prime}$ can then be written as
\begin{eqnarray}
	\mathcal{M}_{i^{\prime}}^{(6)} \mathcal{O}_{i^{\prime}} &=& 
	\left\{ {\rm Tr}(\widehat H_d \widehat H_u \widehat H_d^2 \widehat H_u^2) 
	\mathcal{N}^{i^{\prime}} \right\}
	\widetilde \Phi_2^{\dagger}(p_1) \widetilde \Phi_1(p_2) \widetilde \Phi_2^{\dagger}(p_3) \widetilde \Phi_1(p_4) 
		\nonumber \\
	&=& \left\{ {\rm Tr}(\widehat H_u \widehat H_d \widehat H_u^2 \widehat H_d^2)^* 
	\mathcal{N}^i \right\}
	\Phi_1^{\dagger}(p_2) \Phi_2(p_1) \Phi_1^{\dagger}(p_4) \Phi_2(p_3). 
\end{eqnarray}
The operator involving the external scalars is distinguishable from that of diagram $i$ because the external momenta are assigned differently. However, since this diagram's superficial degree of divergence is zero, all of the local divergences of the six-loop integral are independent of the external momenta. Furthermore, the local divergences are independent of the masses of the internal scalars (this is important because the interchange of $\Phi_1$ and $\widetilde \Phi_2$ would also swap their masses).

\begin{figure}[h!]
 \resizebox{0.24\linewidth}{!}{ \includegraphics{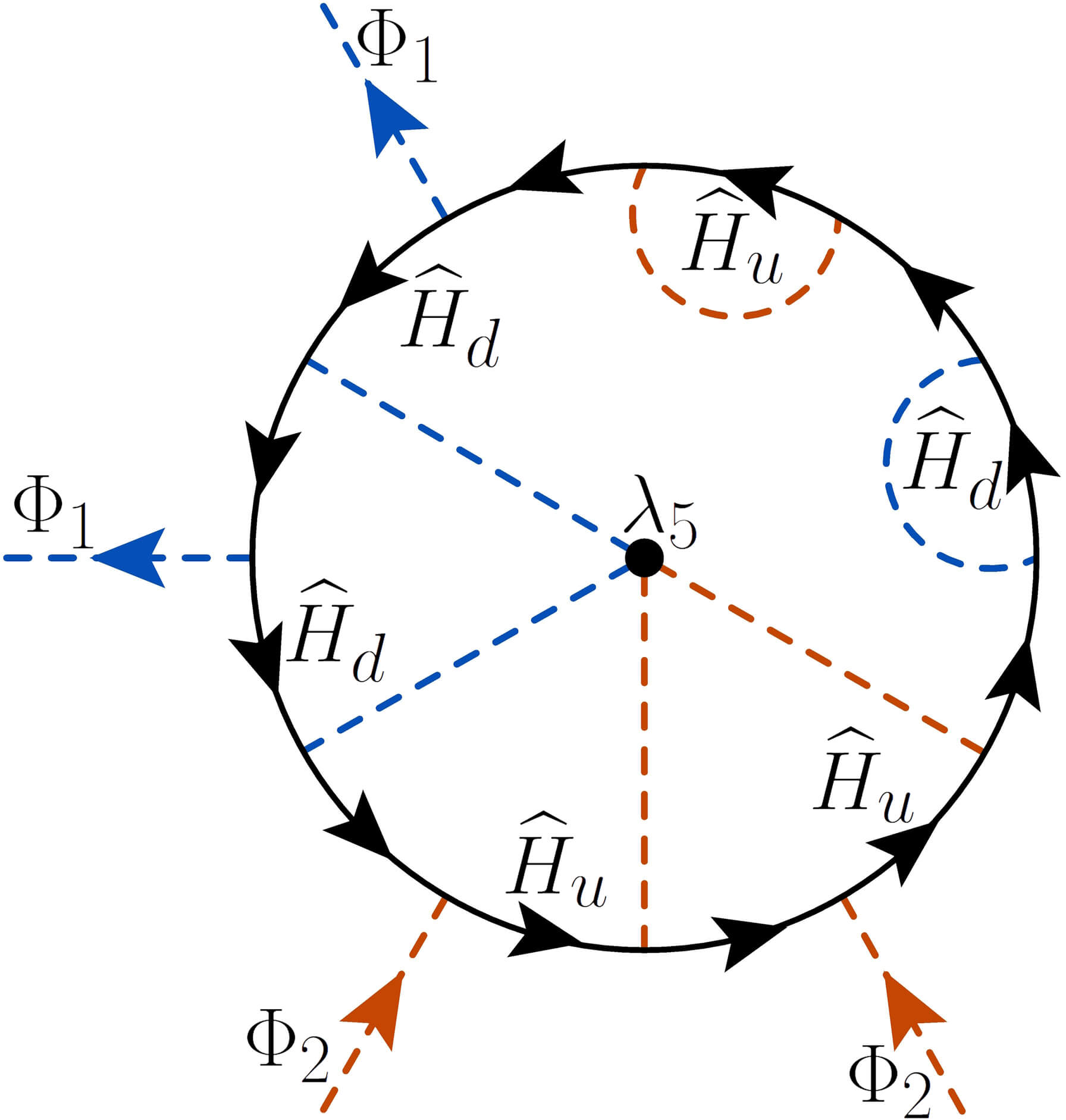}} \, \, \, \, \, \resizebox{0.24\linewidth}{!}{ \includegraphics{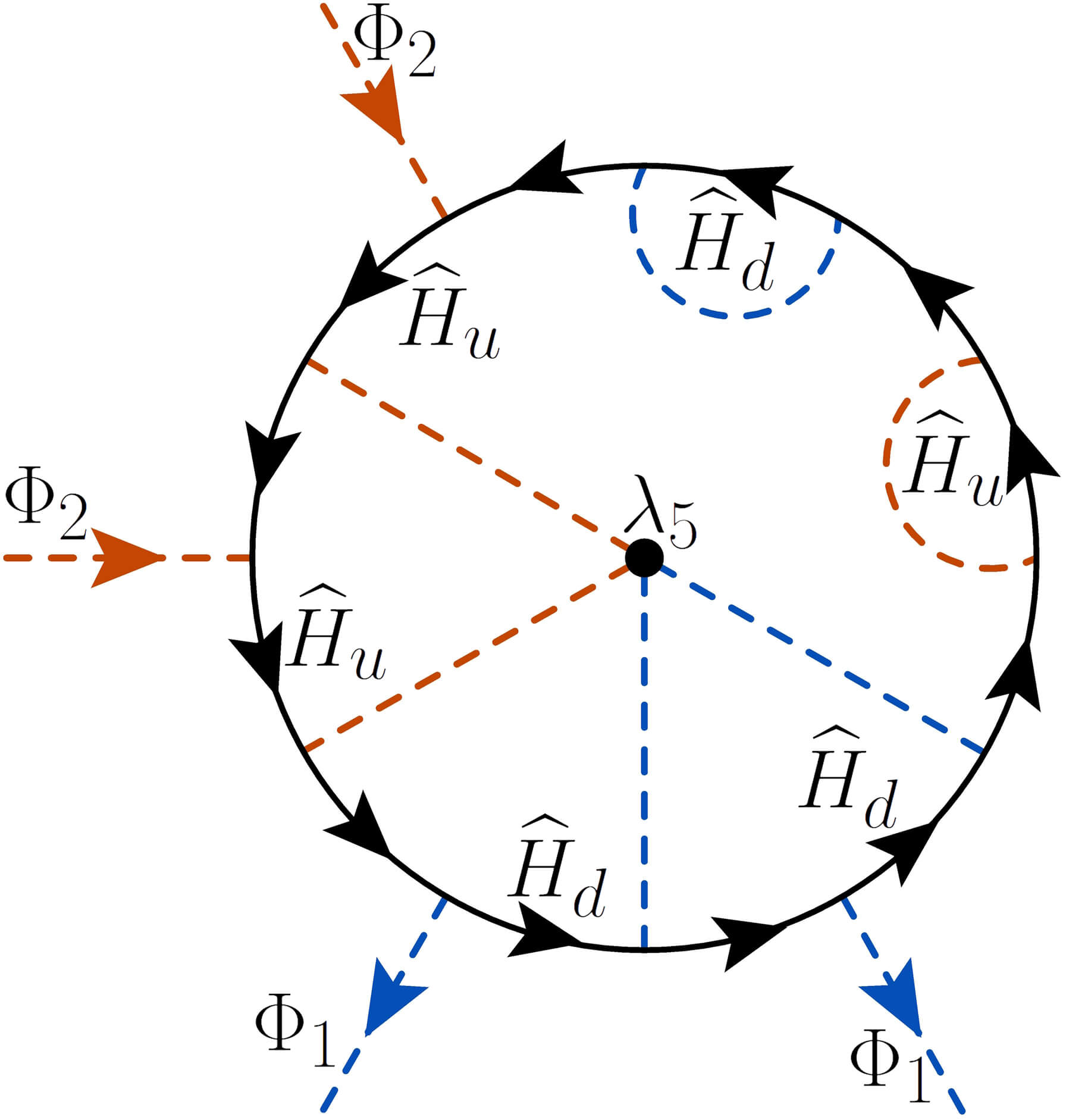}}
\caption{A pair of six-loop diagrams contributing to $\mathcal{O}_5$ in the type II 2HDM, between which the imaginary divergent contributions cancel.}
\label{fig:6loopint}
\end{figure}

We thus show that for the type II real 2DHM at six loops, all divergent pieces of the sum of diagrams are proportional to $\mathcal{J}+\mathcal{J}^{*}$ and thus no imaginary divergence is generated for $\lambda_5$. This cancellation in the type II model can be understood more deeply by observing that the operator $\mathcal{O}_5$ is invariant under the generalized CP transformation $\Phi_1 \to \widetilde \Phi_2$, $\Phi_2 \to \widetilde \Phi_1$. The Yukawa Lagrangian is not invariant under this transformation; however, an invariance can be artificially constructed in the type II model by simultaneously taking $u_R \leftrightarrow d_R$ and $Y_u \leftrightarrow Y_d$.  Interchanging $Y_u \leftrightarrow Y_d$ causes $\mathcal{J} \to \mathcal{J}^*$ (as one would expect from this symmetry's role as a generalized CP transformation), leading to the cancellation of the imaginary part of the six-loop coefficient of $\mathcal{O}_5$ as found above. One such pair of diagrams related by this transformation is shown in Fig.~\ref{fig:6loopint}. We have explicitly verified this cancellation by generating all contributing diagrams using QGRAF~\cite{QGRAF} and identifying the matched pairs.

The realness of $\lambda_5$ is thus protected at six loops by an additional would-be generalized CP symmetry in the type II 2HDM. To break this additional symmetry requires the explicit presence of yet another dimensionless coupling within the contributing Feynman diagrams. This necessitates going to seven loops and is discussed in Sec.~\ref{sec:7loop}.

%%%%%%%%%%%%%%%%%%%%%%%%%%%%%%%%%%%%%%%%%%%
\subsection{Type I 2HDM}
\label{sec:6loopTypeI}
%%%%%%%%%%%%%%%%%%%%%%%%%%%%%%%%%%%%%%%%%%%

We now consider the type I 2HDM. A simple proof of the cancellation of the imaginary divergences as in the previous subsection cannot be made in this model because both of the transformations identified in Sec.~\ref{sec:CPVbase} that take $\mathcal{J} \to \mathcal{J}^*$ also take the operator $\mathcal{O}_5 \to \mathcal{O}_5^{\dagger}$. Instead, we will demonstrate that the imaginary divergent contributions to $\lambda_5$ cancel at six loops in the type I model due to the restricted topologies of all contributing Feynman diagrams.

\begin{figure}[h!]
 \resizebox{0.4\linewidth}{!}{ \includegraphics{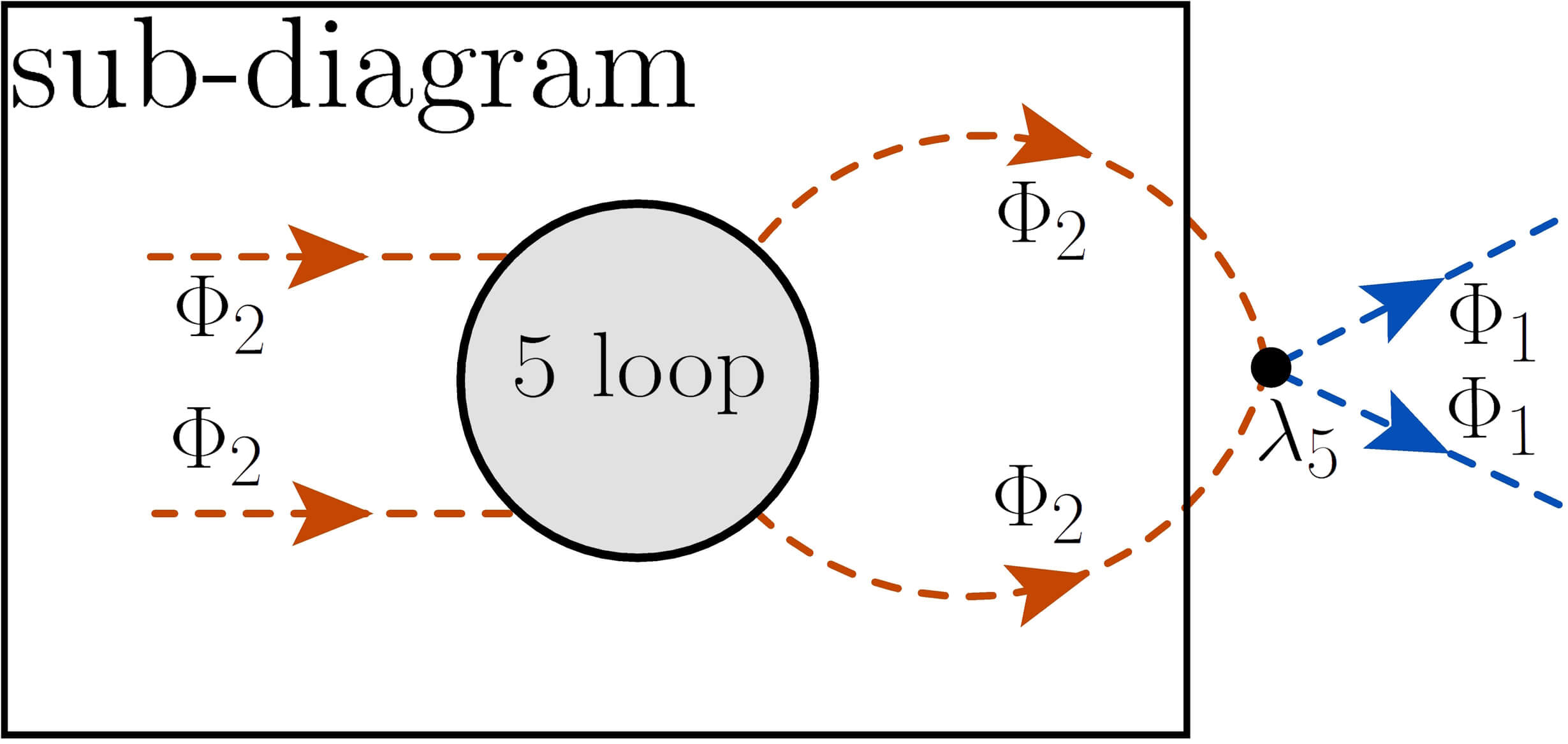}} 
\caption{Subdiagram structure of the six-loop contributions to $\mathcal{O}_5$ in the type I 2HDM.  The subdiagram contains the closed quark loop that yields $\mathcal{J}$ or $\mathcal{J}^*$.}
\label{fig:EFT}
\end{figure}

The key observation is that because in the type I model only $\Phi_2$ couples to quarks, all six-loop diagrams contributing to $\mathcal{O}_5$ that are proportional to $\mathcal{J}$ or $\mathcal{J}^*$ have the structure of a five-loop subdiagram with two legs connected to a $\lambda_5$ vertex, as shown in Fig.~\ref{fig:EFT}. All of the dependence of the six-loop result on $\mathcal{J}$ or $\mathcal{J}^*$ is thus contained within the form factor of the five-loop subdiagram. However, this subdiagram corresponds to the operator $\mathcal{O}_2 \equiv \Phi_2^{\dagger} \Phi_2 \Phi_2^{\dagger} \Phi_2$, which is dimension-four and Hermitian.  This fact will allow us to prove that all relevant parts of the five-loop form factor must be proportional to $\mathcal{J} + \mathcal{J}^*$ and hence real,\footnote{That the divergent parts of the five-loop form factor are purely real can be immediately understood by noticing that they also constitute the five-loop renormalization of the real coupling $\lambda_2$. We address the finite part of the form factor below.} so that the imaginary divergent contribution to $\lambda_5$ vanishes at six loops. 

Consider an individual five-loop subdiagram that is proportional to $\mathcal{J} = {\rm Tr}(\widehat H_u \widehat H_d \widehat H_u^2 \widehat H_d^2)$.  This subdiagram can be written as
\begin{equation}
	\mathcal{M}_i^{(5)} \mathcal{O}_i = 
	\left\{ {\rm Tr}(\widehat H_u \widehat H_d \widehat H_u^2 \widehat H_d^2) 
	{\rm Tr}(\gamma^{\mu_1} \cdots \gamma^{\mu_{12}})
	\mathcal{N}^i_{\mu_1 \cdots \mu_{12}} \right\}
	\Phi_2^{\dagger}(p_1) \Phi_2(p_2) \Phi_2^{\dagger}(p_3) \Phi_2(p_4),
	\label{eq:MiOiTypeI}
\end{equation}
where the $SU(2)_L$ index structure of the doublets has been captured in the contractions of the four external scalar fields. The factor $\mathcal{N}^i_{\mu_1 \cdots \mu_{12}}$ contains the rest of the loop integral and is a function of the four-momenta $p_{1 \cdots 4}$ and the common auxiliary mass introduced to regulate infrared divergences. At this stage, the four-momenta of the external scalars can be off-shell. 

For each such diagram $i$, a second diagram $i^{\prime}$ exists with identical topology and momentum structure but in which the flow of fermion number is reversed. As discussed in Sec.~\ref{sec:CPVbase}, this replaces $\mathcal{J}$ with $\mathcal{J}^*$ and interchanges $\Phi_2 \leftrightarrow \Phi_2^*$.  It also introduces 12 minus signs from $p^{\mu} \to -p^{\mu}$ in the numerators of the 12 fermion propagators (giving an overall positive sign) and reverses the order of the 12 gamma matrices in the Dirac trace (which is equal to the original trace by a familiar identity).  The $SU(2)_L$ index contractions remain unaffected. Because the topology and momentum structure of diagram $i^{\prime}$ is identical to that of diagram $i$, the result of the remainder of the loop integral is the same, i.e., $\mathcal{N}^{i^{\prime}}_{\mu_1 \cdots \mu_{12}} = \mathcal{N}^{i}_{\mu_1 \cdots \mu_{12}}$.
Subdiagram $i^{\prime}$ can then be written as
\begin{eqnarray}
	\mathcal{M}_{i^{\prime}}^{(5)} \mathcal{O}_{i^{\prime}} &=& 
	\left\{ {\rm Tr}(\widehat H_d^2 \widehat H_u^2 \widehat H_d \widehat H_u) 
	(-1)^{12} {\rm Tr}(\gamma^{\mu_{12}} \cdots \gamma^{\mu_1})
	\mathcal{N}^{i^{\prime}}_{\mu_1 \cdots \mu_{12}} \right\}
	\Phi_2^{T}(p_1) \Phi_2^*(p_2) \Phi_2^{T}(p_3) \Phi_2^*(p_4) \nonumber \\
	&=& \left\{ {\rm Tr}(\widehat H_u \widehat H_d \widehat H_u^2 \widehat H_d^2)^* 
	{\rm Tr}(\gamma^{\mu_1} \cdots \gamma^{\mu_{12}})
	\mathcal{N}^i_{\mu_1 \cdots \mu_{12}} \right\}
	\Phi_2^{\dagger}(p_2) \Phi_2(p_1) \Phi_2^{\dagger}(p_4) \Phi_2(p_3),
\end{eqnarray}
where in the last line we have also taken the transpose of the products of scalar doublets.

The operator involving the external scalars is distinguishable from that of diagram $i$ because the momenta are assigned differently. This matters because the matrix element associated with this subdiagram is, in general, a Lorentz-invariant function of the incoming four-momenta $p_1, p_2, p_3, p_4$; in going to subdiagram $i^{\prime}$ we have replaced the original kinematic variables according to $p_1 \leftrightarrow p_2$, $p_3 \leftrightarrow p_4$.  
However, since this subdiagram's superficial degree of divergence is zero, all of the local divergences of the five-loop integral are independent of the external momenta. Therefore the imaginary part of $\mathcal{J}$ multiplying divergent terms in $\mathcal{N}$, as well as any finite terms in $\mathcal{N}$ that is independent of the external momenta, cancels in the sum of diagrams $i$ and $i^{\prime}$. 
This result can also be proved trivially by noticing that the five-loop subdiagram in the limit $p_{1 \cdots 4} \to 0$ is just the five-loop renormalization of $\lambda_2$, which multiplies the Hermitian operator $\mathcal{O}_2$ in the scalar potential and is thus guaranteed to remain real at all orders in perturbation theory.

\begin{figure}[h!]
 \resizebox{0.30\linewidth}{!}{ \includegraphics{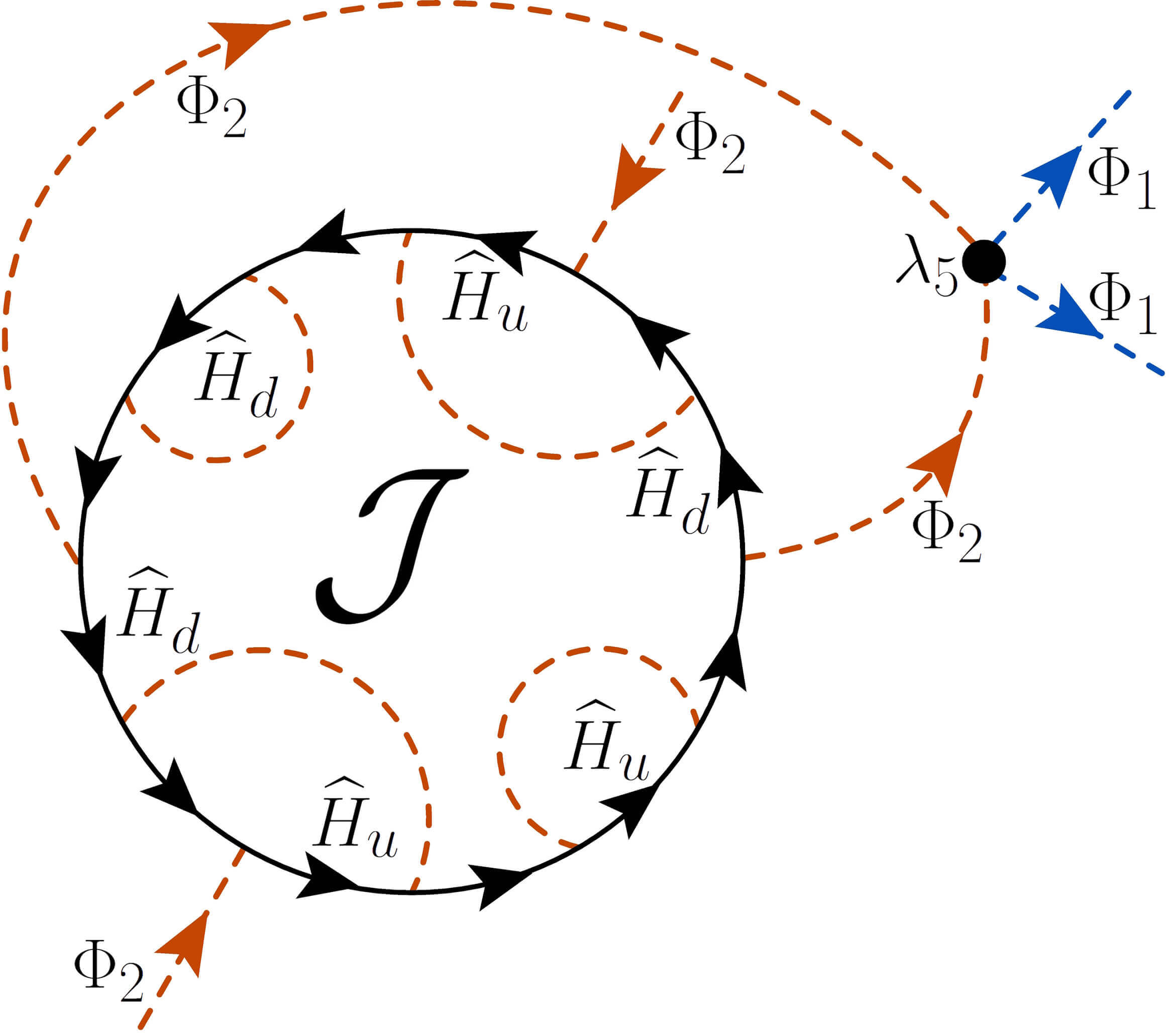}} \, \, \, \, \,   \resizebox{0.33\linewidth}{!}{ \includegraphics{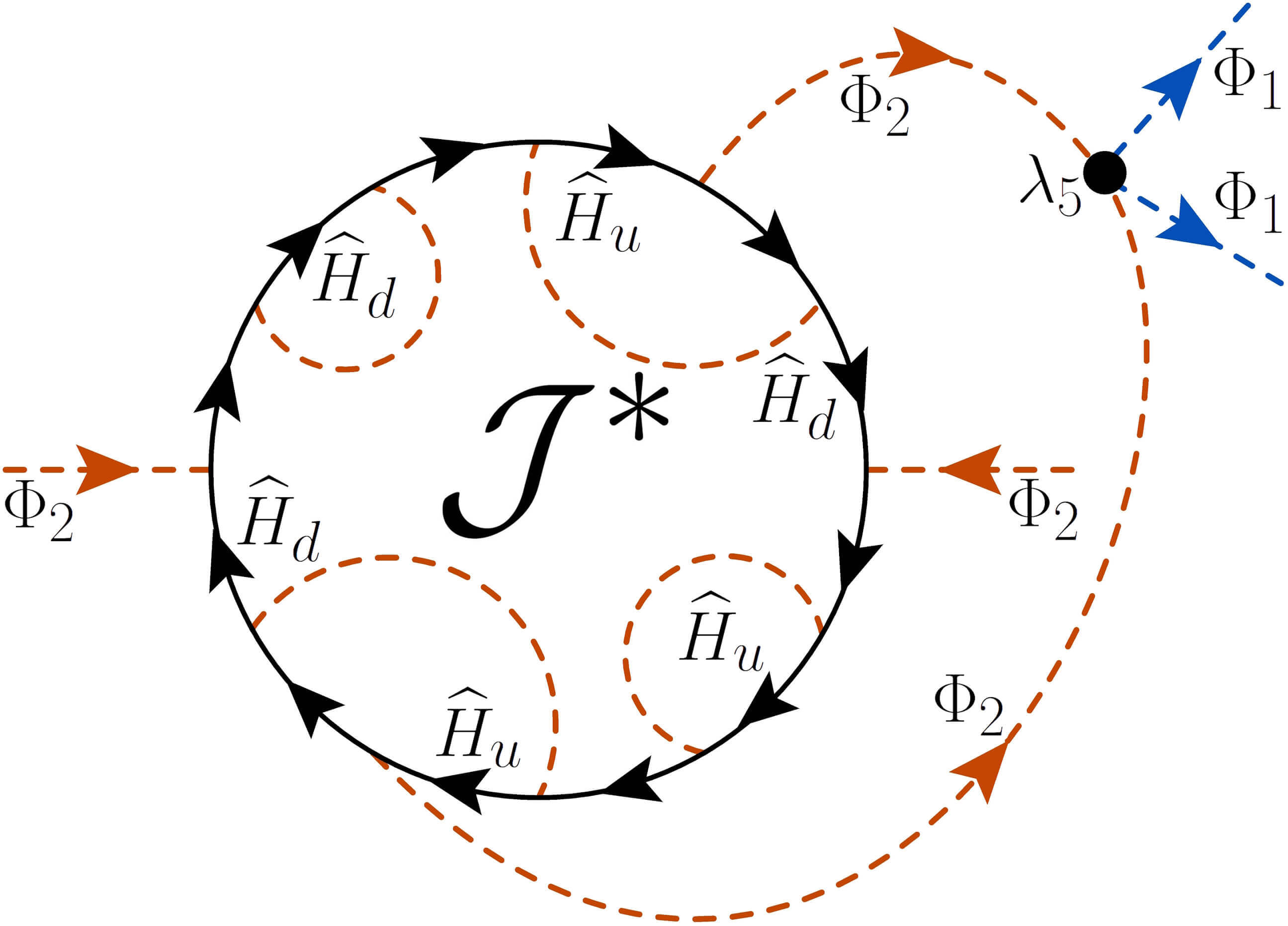}}
\caption{A pair of six-loop diagrams contributing to $\mathcal{O}_5$ in the type I 2HDM with Yukawa-vertex corrections on the two incoming (left) and two outgoing (right) $\Phi_2$ legs of the five-loop subdiagram. Note that the internal closed scalar lines in the second diagram cannot be rearranged to produce a diagram with Yukawa vertex corrections on the incoming legs because of the hypercharge flow.}
\label{fig:eee}
\end{figure}

This leaves only the momentum-dependent terms in $\mathcal{N}$. These matter because the five-loop subdiagram has two off-shell legs (which connect to the $\lambda_5$ vertex, forming the sixth loop), and the reversal of fermion flow switches which pair of legs are off-shell, as shown in Fig.~\ref{fig:eee}. If the momentum dependence of the five-loop subdiagram contains an antisymmetric term under the interchange of the choice of off-shell legs, it could induce a divergent CP-violating contribution at six loops.\footnote{In the effective operator language, this piece of the five-loop form factor corresponds to C-odd operators such as $[\Phi_2^{\dagger} \partial^2 \Phi_2 - (\partial^2 \Phi_2^{\dagger}) \Phi_2] (\Phi_2^{\dagger} \Phi_2)$.}  Fortunately, it is easy to demonstrate that such a term cannot appear once all contributions to the five-loop subdiagram are summed.

In the massless theory, the five-loop form factor is an analytic dimensionless function of Lorentz-invariant combinations of the external four-momenta. Let us write $p_1 = -p_3 = k_a$, $p_2 = -p_4 = k_b$, where $k_a$ and $k_b$ are the momenta of the sixth loop in the left and right diagrams of Fig.~\ref{fig:eee}, respectively. Then the Mandelstam variables entering the five-loop form factor are $s=0$, $t = (k_a + k_b)^2$, and $u = (k_a - k_b)^2$, and any antisymmetric piece generically takes the form $\Lambda(p_i) \sim (k_a^2 - k_b^2)/f(t, u) + \cdots$, where $f(t,u)$ is some linear combination of $t$ and $u$ and the ellipses indicate additional dimensionless terms with higher powers of momenta in both the numerator and denominator.\footnote{Such a structure can arise from individual five-loop diagrams containing one-loop Yukawa vertex corrections on either the two incoming or the two outgoing scalar lines but not both; we show two such diagrams related by reversal of fermion flow in Fig.~\ref{fig:eee}.}  However, in the limit $k_a \to 0$ and $k_b \to 0$, the momentum-dependent piece of the five-loop form factor must reduce to a well-defined, finite limit corresponding to the physical process of massless $\Phi_2 \Phi_2 \to \Phi_2 \Phi_2$ scattering at threshold; furthermore if in this limit the form factor is nonzero, its coefficient cannot be complex because it becomes a contribution to the renormalization of $\lambda_2$. The antisymmetric piece of the form factor described above is not only finite in this limit, but its sign depends on the order in which the limits $k_a \to 0$ and $k_b \to 0$ are taken, which is clearly unphysical. Thus, we prove that the coefficient of any such antisymmetric piece must be zero; i.e., that the momentum-dependent piece of the five-loop form factor must be symmetric under the interchange of the incoming and outgoing legs. Therefore, the entire form factor must be proportional to $\mathcal{J} + \mathcal{J}^*$ and hence real.  

Embedding this form factor into the sixth loop, we thus demonstrate that all imaginary divergent contributions to $\lambda_5$ cancel at six loops in the type I model. We emphasize that this cancellation at six loops is a diagrammatic accident in the type I model and is not protected by any symmetry that we have identified; instead, it is due to all contributing diagrams having the subdiagram structure shown in Fig.~\ref{fig:EFT}.

%%%%%%%%%%%%%%%%%%%%%%%%%%%%%%%%%%%%%%%%%%%
\section{$\lambda_5$ at seven loops}
\label{sec:7loop}
%%%%%%%%%%%%%%%%%%%%%%%%%%%%%%%%%%%%%%%%%%%

We showed in the previous section that the imaginary divergent contributions to $\lambda_5$ are zero at the six-loop level in both the type II and type I 2HDMs. We now extend our analysis to seven loops and show that the arguments used to demonstrate the cancellation of the six-loop diagrams no longer hold, so that an imaginary divergent contribution to $\lambda_5$ can arise at this order. We explicitly identify the classes of diagrams that can contribute and the resulting parameter dependence that can appear in the seven-loop RG equations, thereby laying the groundwork for future explicit calculations. Again, we analyze the two types separately.

Since we have not performed the loop integrals to calculate the divergent parts of the contributing diagrams, we cannot exclude the possibility that some as-yet-unidentified symmetry among diagrams leads to cancellations of the imaginary divergent contributions to $\lambda_5$ also at the seven-loop order.

%%%%%%%%%%%%%%%%%%%%%%%%%%%%%%%%%%%%%%%%%%%
\subsection{Type II 2HDM}
%%%%%%%%%%%%%%%%%%%%%%%%%%%%%%%%%%%%%%%%%%%

In Sec.~\ref{sec:6loopTypeII} we demonstrated that the cancellation of the imaginary divergent contribution to $\lambda_5$ at six loops in the type II 2HDM was guaranteed by the transformation properties of the contributing diagrams under the generalized CP transformation $\Phi_1 \leftrightarrow \widetilde \Phi_2$ along with $u_R \leftrightarrow d_R$ and $Y_u \leftrightarrow Y_d$. In particular, for each six-loop diagram $i$ proportional to $\lambda_5 \mathcal{J}$, this transformation yielded a second diagram $i^{\prime}$ with identical momentum structure proportional to $\lambda_5 \mathcal{J}^*$, between which the imaginary part of $\mathcal{J}$ canceled. The obvious way to destroy this cancellation and potentially recover an imaginary divergent contribution to $\lambda_5$ at the seven-loop order is by introducing an additional coupling multiplying $\lambda_5 \mathcal{J}$ in diagram $i$ that is not the same in diagram $i^{\prime}$. We identify three possible ways to do this:

\begin{enumerate}

\item Insert an additional quartic scalar interaction involving the coupling $\lambda_1$ or $\lambda_2$. Under the generalized CP transformation $\Phi_1 \leftrightarrow \widetilde \Phi_2$, the most general scalar potential of the 2HDM given in Eq.~(\ref{eq:scalarpot}) transforms according to $\lambda_1 \leftrightarrow \lambda_2$, $\lambda_6 \leftrightarrow \lambda_7^*$, and $m_{11}^2 \leftrightarrow m_{22}^2$, with all other terms invariant.  Thus a seven-loop diagram $i$ proportional to $\lambda_1 \lambda_5 \mathcal{J}$ contributing to $\mathcal{O}_5$ transforms into a diagram $i^{\prime}$ proportional to $\lambda_2 \lambda_5 \mathcal{J}^*$, and the cancellation of the imaginary part of $\mathcal{J}$ is spoiled when $\lambda_1 \neq \lambda_2$.  A pair of such seven-loop diagrams is shown in Fig.~\ref{fig:7loopdi2}.  The resulting imaginary divergent contribution to $\lambda_5$ will be proportional to $(\lambda_1 - \lambda_2) \lambda_5 {\rm Im}(\mathcal{J})$.

\begin{figure}[h!]
 \resizebox{0.24\linewidth}{!}{ \includegraphics{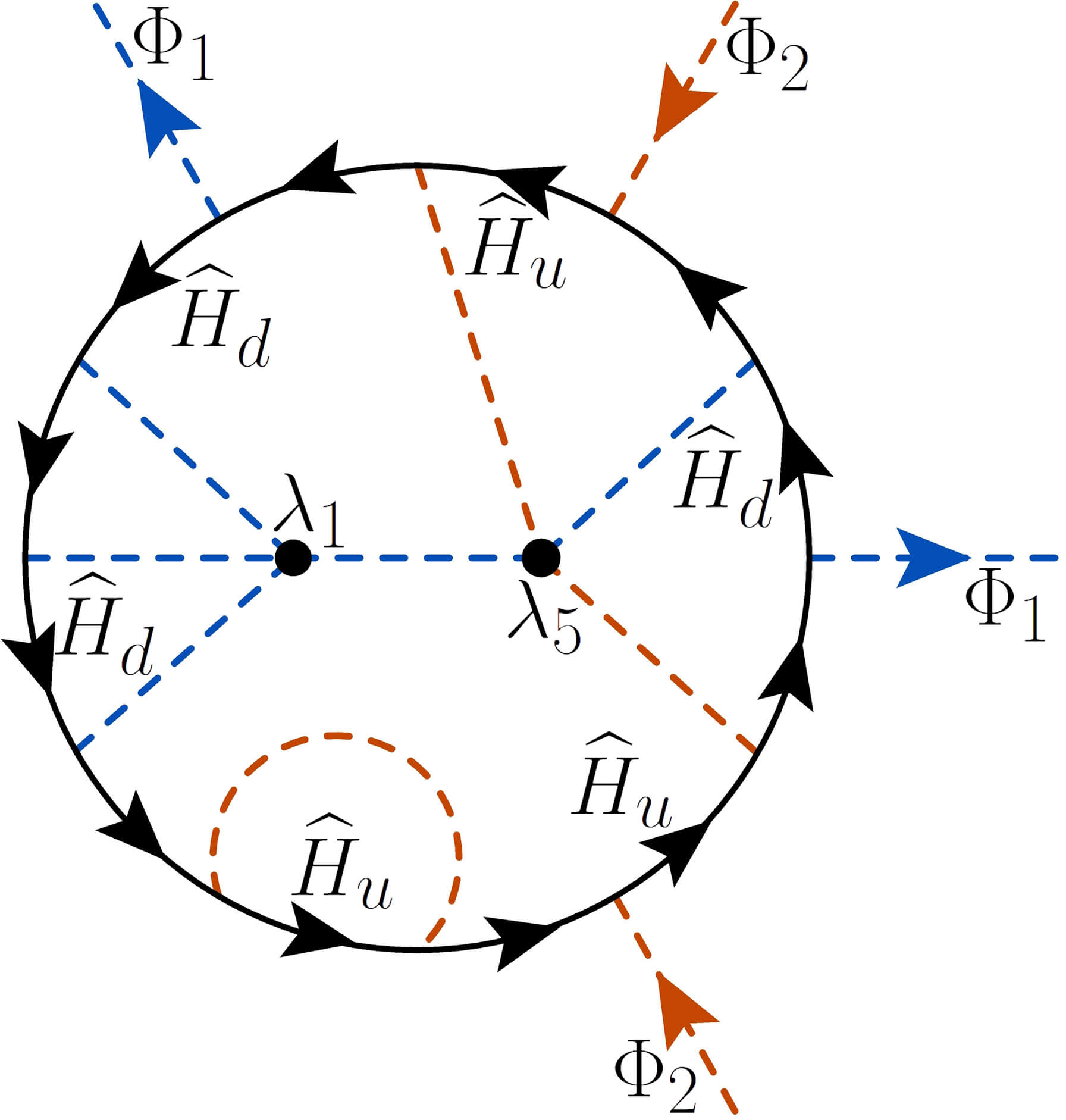}}  \, \, \, \, \, \,  \resizebox{0.24\linewidth}{!}{ \includegraphics{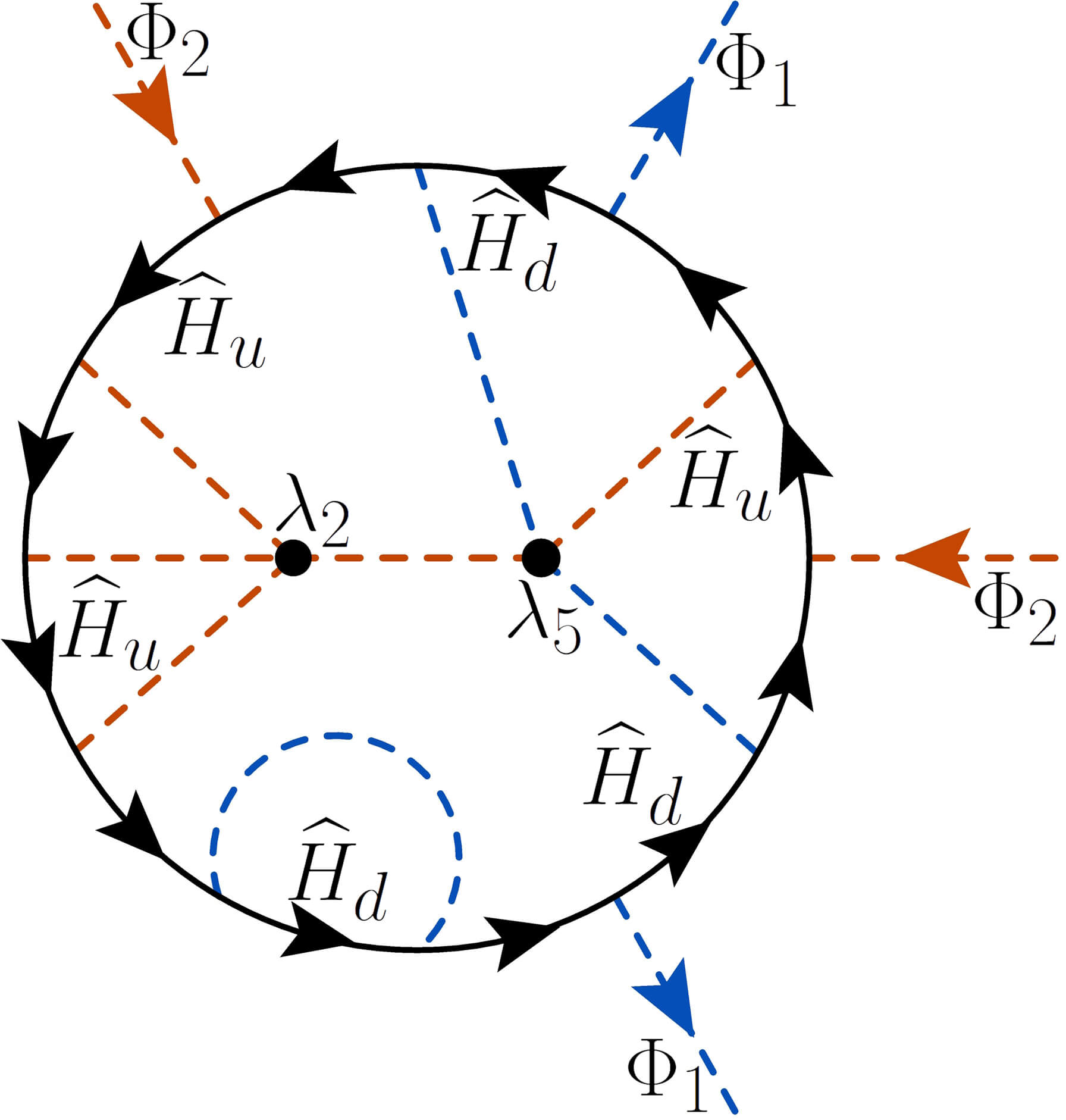}}
\caption{A pair of seven-loop diagrams in the type II 2HDM proportional to $\lambda_1 \lambda_5 \mathcal{J}$ and $\lambda_2 \lambda_5 \mathcal{J}^*$, respectively. There is no possible diagram proportional to $\lambda_1 \lambda_5 \mathcal{J}^{*}$ ($\lambda_2 \lambda_5 \mathcal{J}$) that has the same topology as the diagram on the left (right).}
\label{fig:7loopdi2}
\end{figure}

\item Insert an appropriate hypercharge interaction. The pairs of diagrams between which the imaginary part of $\mathcal{J}$ do not cancel are those in which the product of the two hypercharge quantum numbers involved is not invariant under the transformation $u_R \leftrightarrow d_R$, $\Phi_1 \leftrightarrow \widetilde \Phi_2$. The noninvariant combinations are $(Q_{u_R}^Y)^2 \leftrightarrow (Q_{d_R}^Y)^2$, $Q_{u_R}^Y Q_{Q_L}^Y \leftrightarrow Q_{d_R}^Y Q_{Q_L}^Y$, and $Q_{q}^Y Q_{\Phi}^Y \leftrightarrow -Q_{q}^Y Q_{\Phi}^Y$, where $q$ is any quark and we use the fact that $\widetilde \Phi$ has the opposite hypercharge as $\Phi$.\footnote{In our conventions the hypercharge quantum numbers of the various fields are $Q^Y_{u_R} = +2/3$, $Q^Y_{d_R} = -1/3$, $Q^Y_{Q_L} = 1/6$, and $Q^Y_{\Phi} = 1/2$.} 
A pair of such seven-loop diagrams is shown in Fig.~\ref{fig:pic}. 
The resulting imaginary divergent contributions to $\lambda_5$ will be proportional to $g^{\prime 2} \lambda_5 {\rm Im}(\mathcal{J})$ times the differences $Q_i^YQ_j^Y - \widetilde Q_i^Y \widetilde Q_j^Y$ of the appropriate pairs of hypercharge quantum numbers.

\begin{figure}[h!]
 \resizebox{0.24\linewidth}{!}{ \includegraphics{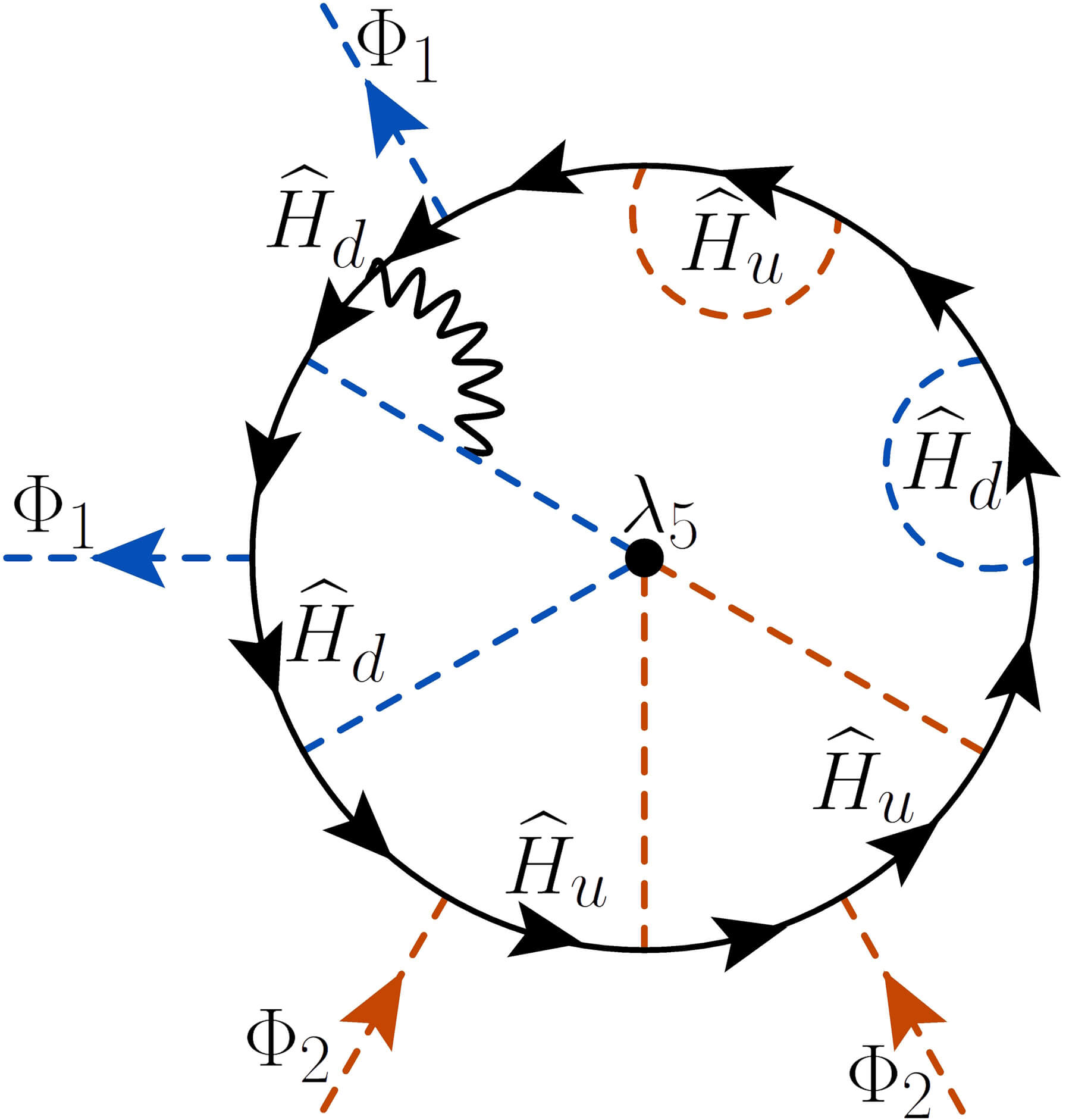}}  \, \, \, \, \, \,  \resizebox{0.24\linewidth}{!}{ \includegraphics{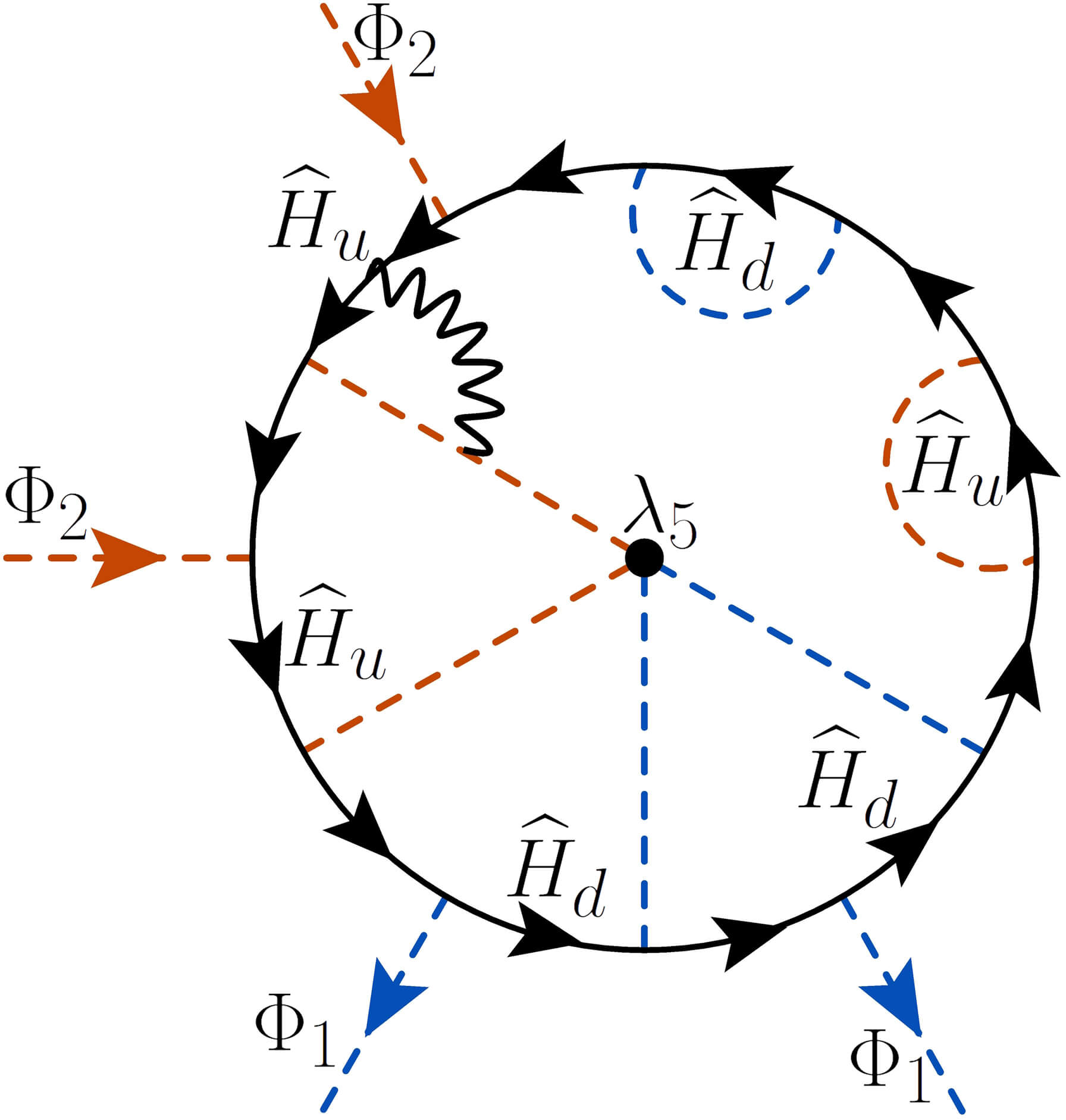}}
\caption{A pair of seven-loop diagrams in the type II 2HDM involving a hypercharge gauge boson exchange between a right-handed quark and a scalar.}
\label{fig:pic}
\end{figure}

\item Insert an additional pair of up-type or down-type Yukawa couplings at an appropriate location in the closed quark loop. Inserting two additional up-type Yukawa couplings yields a Yukawa structure $\mathcal{J}^u \equiv {\rm Tr}(\widehat H_u \widehat H_d \widehat H_u^3 \widehat H_d^2)$, while introducing two additional down-type Yukawa couplings yields $\mathcal{J}^d \equiv {\rm Tr}(\widehat H_u \widehat H_d \widehat H_u^2 \widehat H_d^3)$.\footnote{Inserting the additional Yukawa matrices in different locations than these yields traces that are purely real.}  The imaginary parts of these traces are both proportional to the Jarlskog invariant as in Eq.~(\ref{eq:ImJ}), but with the fourth powers of masses in $T(M_U^2)$ or $B(M_D^2)$ (respectively) in Eq.~(\ref{eq:TB}) replaced with sixth powers along with the corresponding additional factors of $v$ and $\sin\beta$ or $\cos\beta$ in the denominator. Under the transformation $u_R \leftrightarrow d_R$, $\mathcal{J}^u \leftrightarrow \mathcal{J}^{d*}$, so that a seven-loop diagram $i$ proportional to $\lambda_5 \mathcal{J}^u$ contributing to $\mathcal{O}_5$ transforms into a diagram $i^{\prime}$ proportional to $\lambda_5 \mathcal{J}^{d*}$, and the cancellation of the imaginary part is spoiled.  A seven-loop diagram proportional to $\lambda_5 \mathcal{J}^d$ is shown on the left side of Fig.~\ref{fig:pic2}. The resulting imaginary divergent contribution to $\lambda_5$ will be proportional to $\lambda_5 {\rm Im}(\mathcal{J}^{u}-\mathcal{J}^{d}) = (y_t^2 - y_b^2 + \ldots ) \lambda_5 {\rm Im}(\mathcal{J})$, where $y_t = \sqrt{2} m_t / v \sin\beta$ and $y_b = \sqrt{2} m_b / v \cos\beta$ are the top and bottom quark Yukawa couplings. The ellipses denote extra pieces suppressed by second- and first-generation Yukawa couplings, which we neglect here.

\begin{figure}[h!]
 \resizebox{0.27\linewidth}{!}{ \includegraphics{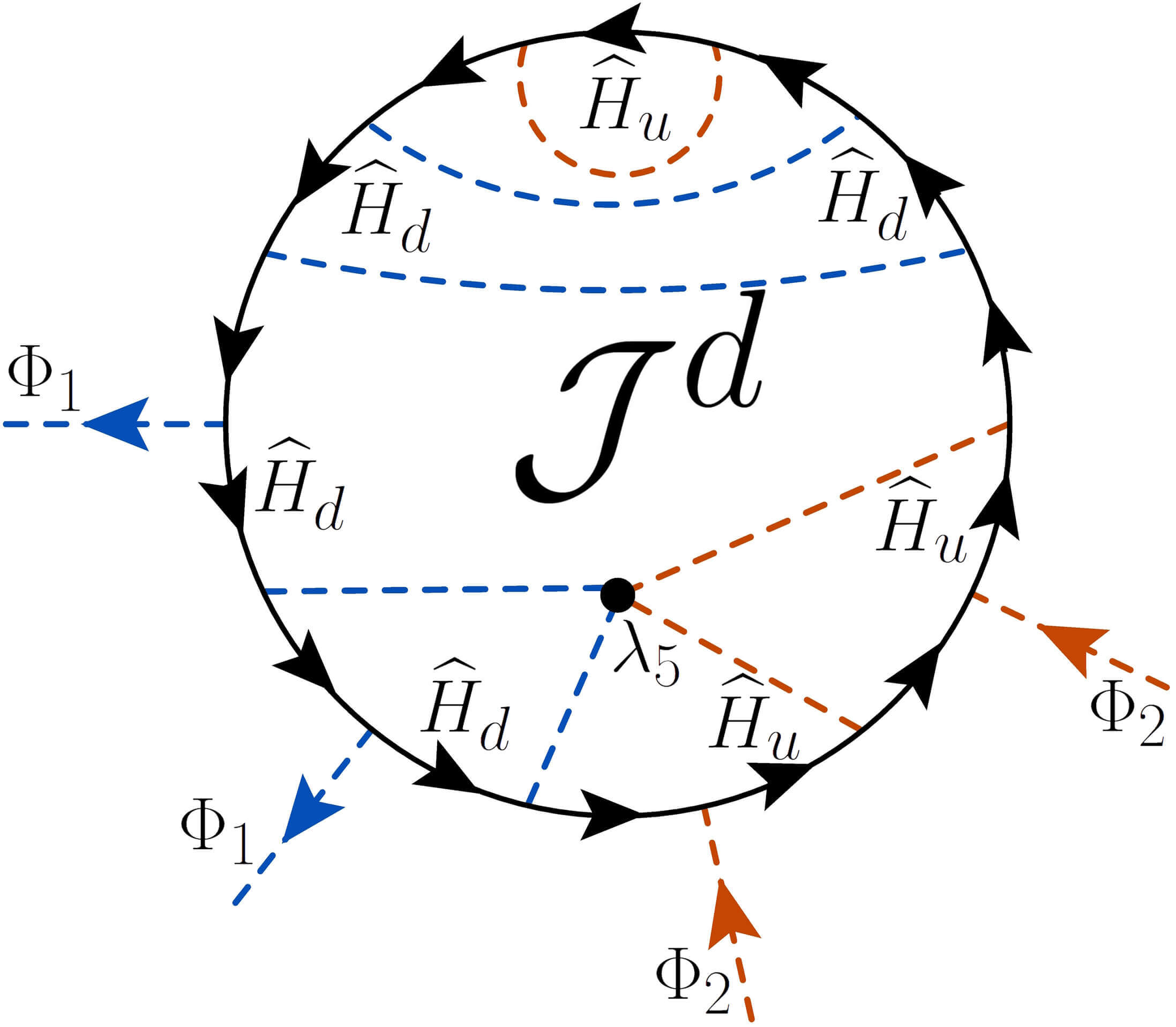}}  \, \, \, \, \,  \resizebox{0.27\linewidth}{!}{ \includegraphics{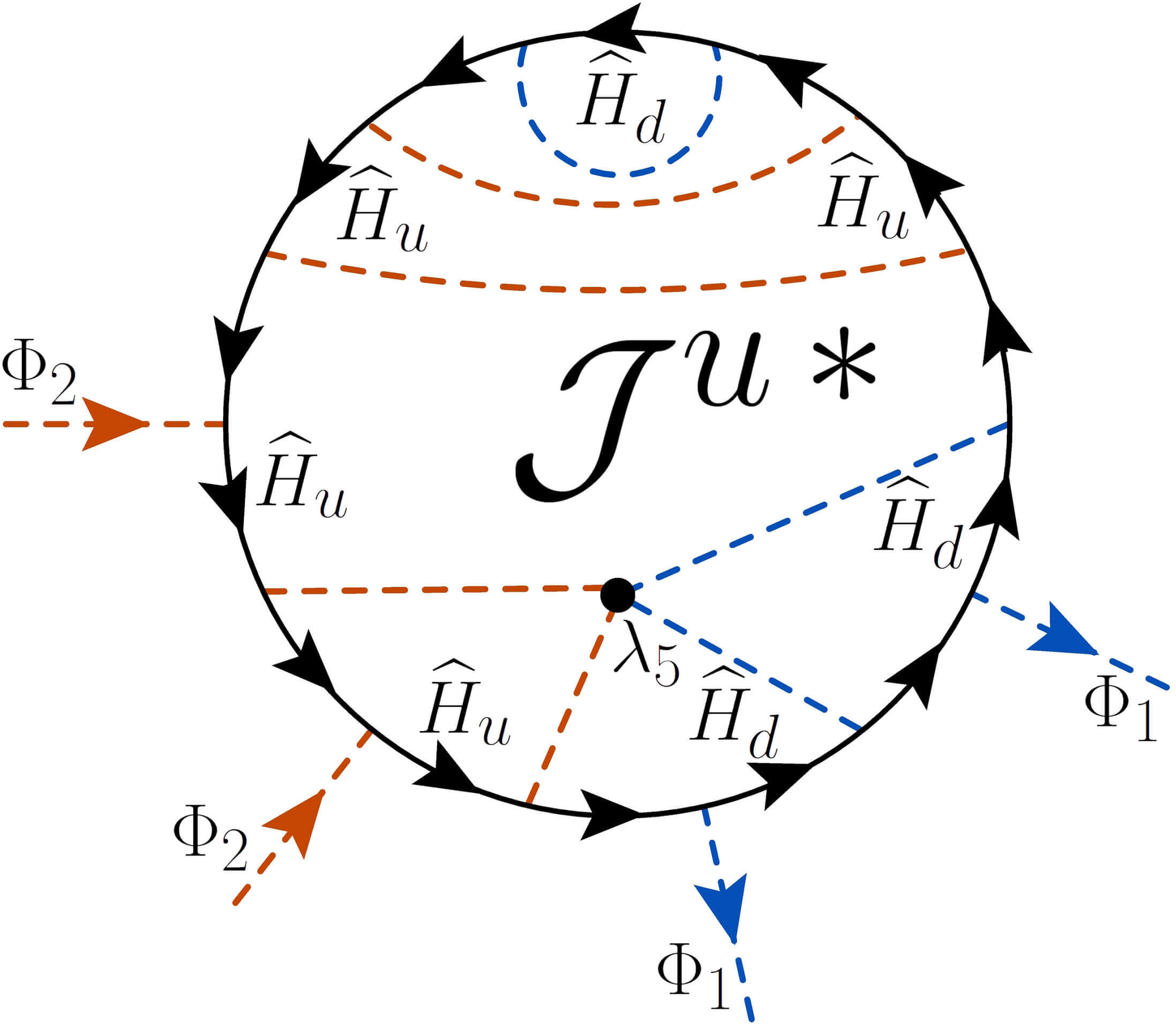}}
\caption{Left: a seven-loop diagram in the type II 2HDM involving a closed quark loop with a total of 14 Yukawa insertions, yielding a Yukawa trace $\mathcal{J}^d \equiv {\rm Tr}(\widehat H_u \widehat H_d \widehat H_u^2 \widehat H_d^3)$. Right: the corresponding diagram with the same topology obtained by interchanging $u_R \leftrightarrow d_R$, which is proportional to $\mathcal{J}^{u*}$, not $\mathcal{J}^{d*}$.}
\label{fig:pic2}
\end{figure}

\end{enumerate}

We thus expect that the leading contribution to the RG equation for ${\rm Im}(\lambda_5)$ in the type II 2HDM can first appear at seven loops and must take the form
\begin{equation}
	\frac{d \, {\rm Im}(\lambda_{5})}{d \ln \mu} 
	= \frac{\lambda_5 {\rm Im}(\mathcal{J})}{(16 \pi^2)^7} \left[
		a^{\lambda} (\lambda_1 - \lambda_2) 
			+ a^{g^{\prime}} g^{\prime 2} 
			+ a^y (y_t^2 - y_b^2 + \ldots ) \right] ,
\end{equation}
where $a^{\lambda}$, $a^{g^{\prime}}$, and $a^y$ are as-yet-uncalculated numerical coefficients, and in the last term, we have neglected additional pieces proportional to second- and first-generation Yukawa couplings.  

Note that each of these contributions would be zero if the additional coupling(s) involved in each diagram proportional to $\mathcal{J}$ were the same as those in the corresponding diagram proportional to $\mathcal{J}^*$.
The symmetries that would lead these contributions to cancel are absent in the type II 2HDM. The second term cannot be zero by symmetry arguments, as this would only be possible if left- and right-handed quarks had the same hypercharge. The first and third terms could initially be set to zero by choosing $\lambda_1 = \lambda_2$ and tuning $\tan\beta$ such that $y_t = y_b$. However, these choices are unstable under the RG running already at one loop~\cite{PhysRevD.56.5366,Fontes:2021iue,Oredsson:2019mni,Oredsson:2018yho,Herren:2017uxn,Bednyakov:2018cmx,Chowdhury:2015yja}. In the absence of a more subtle cancellation between diagrams with different topologies; we expect the numerical coefficients in each term to be nonzero, in which case the real type II 2HDM is theoretically inconsistent.

%%%%%%%%%%%%%%%%%%%%%%%%%%%%%%%%%%%%%%%%%%%
\subsection{Type I 2HDM}
%%%%%%%%%%%%%%%%%%%%%%%%%%%%%%%%%%%%%%%%%%%

In Sec.~\ref{sec:6loopTypeI} we demonstrated that the cancellation of the imaginary divergent contributions to $\lambda_5$ at six loops in the type I 2HDM was guaranteed by the subdiagram structure of all contributing diagrams. In particular, the five-loop subdiagram corresponded to a dimension-four Hermitian operator, which allowed us to prove that all of the relevant pieces of the sum of contributing five-loop diagrams must be real. With this in mind, we consider dressing the six-loop diagrams contributing to $\mathcal{O}_5$ with an additional loop. Three classes of topologies can arise.  

The first class comprises diagrams in which the seventh loop is entirely contained within the Hermitian subdiagram. Examples include inserting an additional quartic scalar interaction within the subdiagram, inserting a gauge boson exchange between propagators of the subdiagram, or inserting an additional pair of Yukawa couplings in the closed quark loop. Since none of these modifications changes the subdiagram structure, they cannot give rise to an imaginary divergent contribution to $\lambda_5$.  

The second class comprises diagrams in which the seventh loop dresses the $\lambda_5$ vertex outside the subdiagram. These diagrams involve a one-loop renormalization of $ \lambda_5 $ (which cannot itself introduce an imaginary part) prior to its attachment to the five-loop subdiagram. Since the cancellation of the relevant imaginary parts of the five-loop subdiagram remains untouched by the seventh loop, this class of diagrams cannot give rise to an imaginary divergent contribution to $\lambda_5$. 

These first two classes yield an immediate difference compared to the type II model in that there are no imaginary divergent contributions in type I involving $\lambda_{1}$ (which would fall into the second class) and none involving $\lambda_{2}$, $\mathcal{J}^{u}$, or $\mathcal{J}^{d}$ (which would fall into the first class).

The third class comprises diagrams in which the seventh loop attaches to both the subdiagram and one of the outgoing $\Phi_1$ legs of the external $\lambda_5$ vertex, thereby fundamentally changing the subdiagram structure. The interactions that can give rise to such a diagram are the exchange of a hypercharge or $SU(2)_L$ gauge boson or the coupling of a $\Phi_2 \Phi_2^*$ pair pulled from the subdiagram to one of the outgoing $\Phi_1$ lines via a $\lambda_3$ or $\lambda_4$ quartic scalar coupling. These diagrams cannot be reduced to the subdiagram structure of Fig.~\ref{fig:EFT} and hence are not subject to our proof of the cancellation of the imaginary divergent parts; thus, they can, in principle, give rise to imaginary divergent contributions to $\lambda_5$. 

\begin{figure}[h!]
 \resizebox{0.3\linewidth}{!}{ \includegraphics{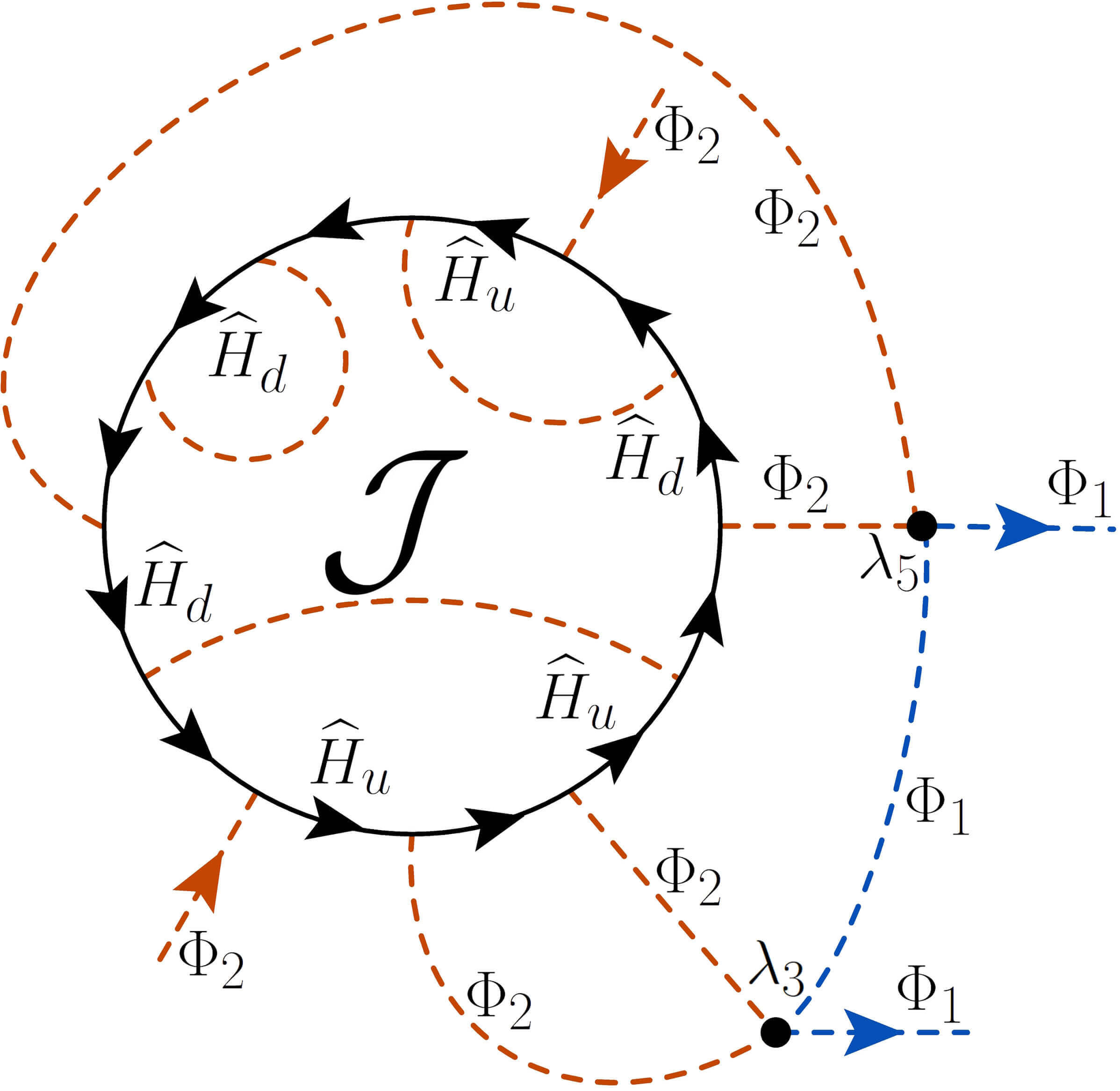}} \, \, \, \,  \resizebox{0.32\linewidth}{!}{ \includegraphics{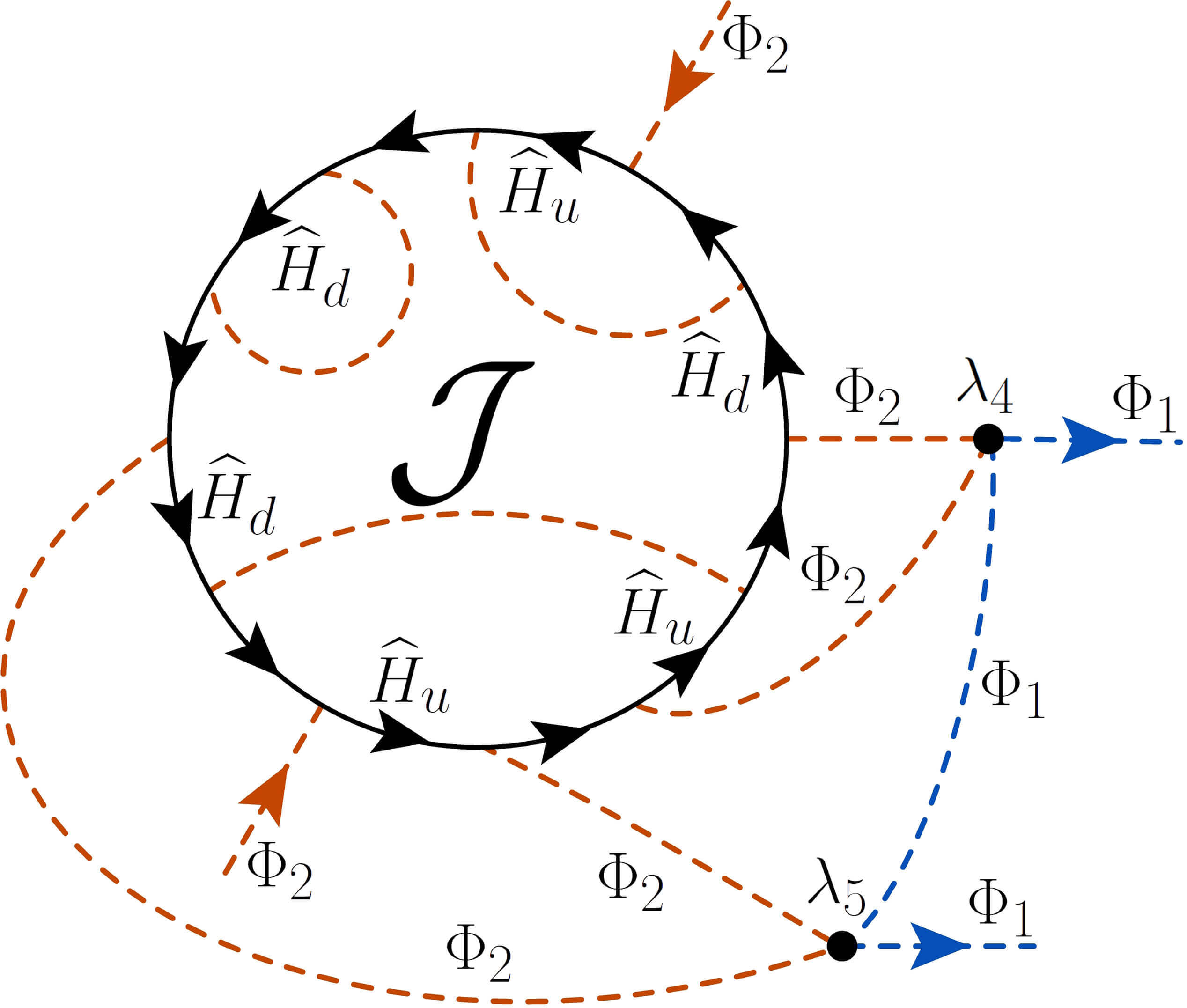}} 
\caption{Seven-loop diagrams in the type I 2HDM involving a $\lambda_{3}$ (left) or $\lambda_{4}$ (right) insertion.}
\label{fig:pic4}
\end{figure}

The topology of the diagrams involving $\lambda_{3}$ or $\lambda_{4}$ is restricted by the $SU(2)_{L}$ structure of the corresponding four-scalar vertices.  In particular, a diagram involving a $\lambda_{3}$ insertion cannot be transformed into a diagram involving $\lambda_4$ simply by swapping these two vertices. We have not identified any deeper relationship between the diagrams involving $\lambda_{3}$ or $\lambda_{4}$ and have not identified any reason for cancellations of the imaginary parts between pairs of diagrams in this case.  Two such diagrams are shown in Fig.~\ref{fig:pic4}.
 
\begin{figure}[h!]
 \resizebox{0.30\linewidth}{!}{ \includegraphics{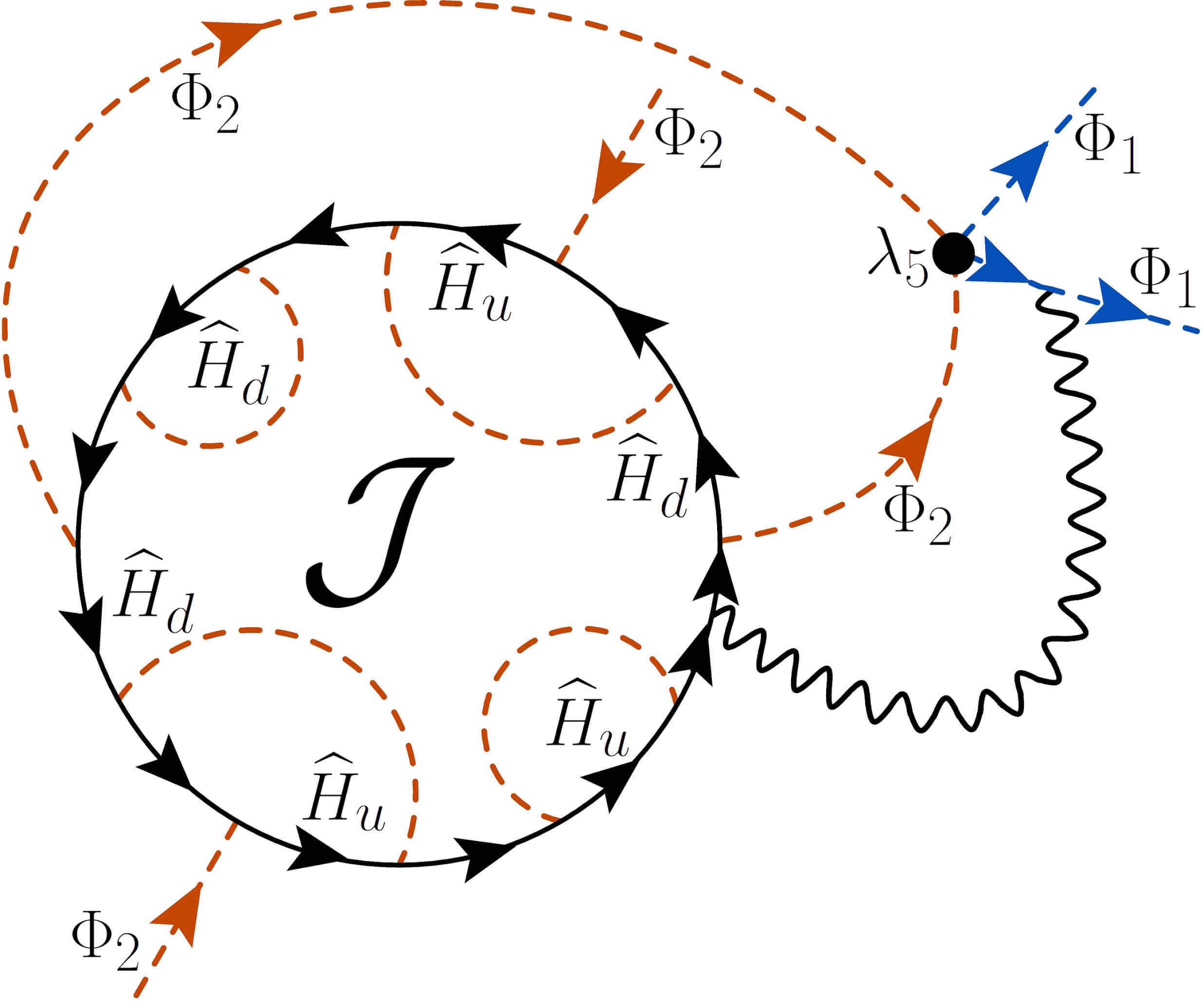}} \,  \,  \resizebox{0.33\linewidth}{!}{ \includegraphics{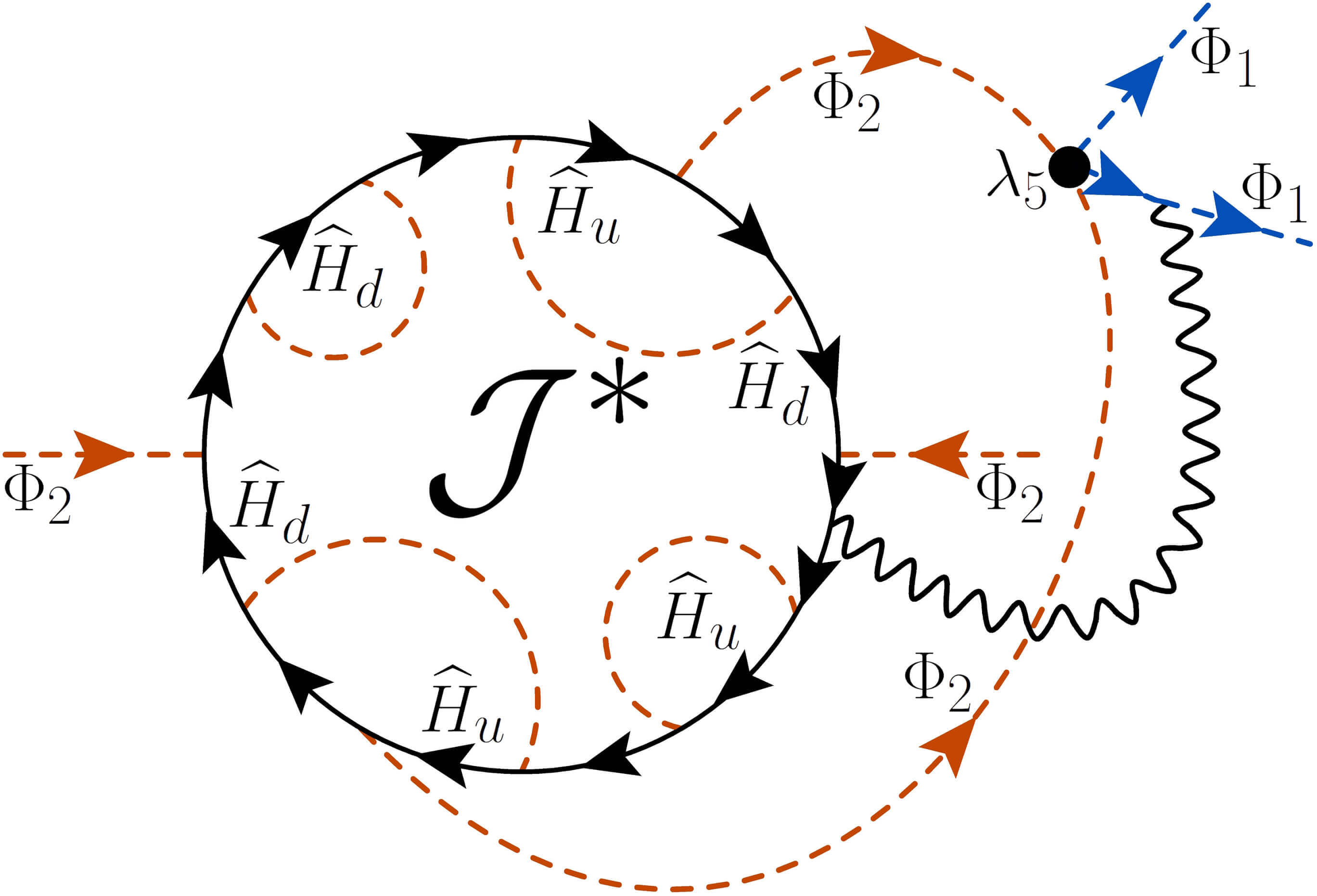}}
\caption{A pair of seven-loop diagrams in the type I 2HDM involving a gauge boson exchange.}
\label{fig:pic3}
\end{figure}

The non-cancellation of the imaginary divergent piece of the diagrams with a hypercharge or $SU(2)_L$ gauge boson exchange is perhaps less clear. Two such diagrams, related by reversing the flow of fermion number and swapping which pair of scalar legs is connected to the $\lambda_5$ vertex, are shown in Fig.~\ref{fig:pic3}. One could imagine reducing these diagrams to a subdiagram structure by performing six of the seven loop integrals, leaving only the final gauge boson and $\Phi_1$ loop; however, the resulting form factor now has mass-dimension $-2$, so our proof in the previous section relying on its reduction to a physical two-to-two scattering process no longer holds. An alternative line of argument could be constructed based on the fact that once the Yukawa coupling structure is factored out of the Feynman diagram, all the remaining propagators in the primitive $\mathcal{J}$ and $\mathcal{J}^{*}$ diagrams are indistinguishable under $u_R \leftrightarrow d_R$, which seems to indicate that the gauge interactions should not be able to distinguish these primitive diagrams; however, we have not been able to construct a conclusive proof along these lines.

We thus expect that the leading contribution to the RG equation for ${\rm Im}(\lambda_5)$ in the type I 2HDM can first appear at seven loops and must take the form
\begin{equation}
	\frac{d \, {\rm Im}(\lambda_{5})}{d \ln \mu} 
	= \frac{\lambda_5 {\rm Im}(\mathcal{J})}{(16 \pi^2)^7} \left[
		b^{\lambda_3} \lambda_3 + b^{\lambda_4} \lambda_4 + b^{g^{\prime}} g^{\prime 2} + b^g g^2
		 \right] ,
\end{equation}
where $b^{\lambda_3}$, $b^{\lambda_4}$, $b^{g^{\prime}}$, and $b^g$ are as-yet-uncalculated numerical coefficients.  In contrast to the type II model, no would-be symmetry-restoring relationship exists between the couplings $\lambda_3$ and $\lambda_4$ that would lead the seven-loop contribution proportional to these couplings to vanish. Again, in the absence of a more subtle cancellation between diagrams with different topologies, we expect the numerical coefficients in each term to be nonzero, in which case the real type I 2HDM is theoretically inconsistent.

%%%%%%%%%%%%%%%%%%%%%%%%%%%%%%%%%%%%%%%%%%%
\section{Corrections to $m_{12}^2$}
\label{sec:m12sq}
%%%%%%%%%%%%%%%%%%%%%%%%%%%%%%%%%%%%%%%%%%%

We finally consider the generation of imaginary divergent contributions to $m_{12}^2$. As discussed in Sec.~\ref{sec:symcons}, any such contribution to $m_{12}^2$ must be proportional to at least $m_{12}^2 {\rm Im}(\mathcal{J})$ or $m_{12}^{2*} \lambda_5 {\rm Im}(\mathcal{J})$.  Since the RG equation for the dimension-two parameter $m_{12}^2$ must already involve one power of $m_{12}^2$ (or its complex conjugate), the other mass-squared parameters $m_{11}^2$ and $m_{22}^2$ cannot appear in the RG equation for $m_{12}^2$ on dimensional grounds. We can thus continue to treat the doublets $\Phi_1$ and $\Phi_2$ as massless scalar degrees of freedom and treat $m_{12}^2$ as a mass insertion.

\begin{figure}[h!]
 \resizebox{0.33\linewidth}{!}{ \includegraphics{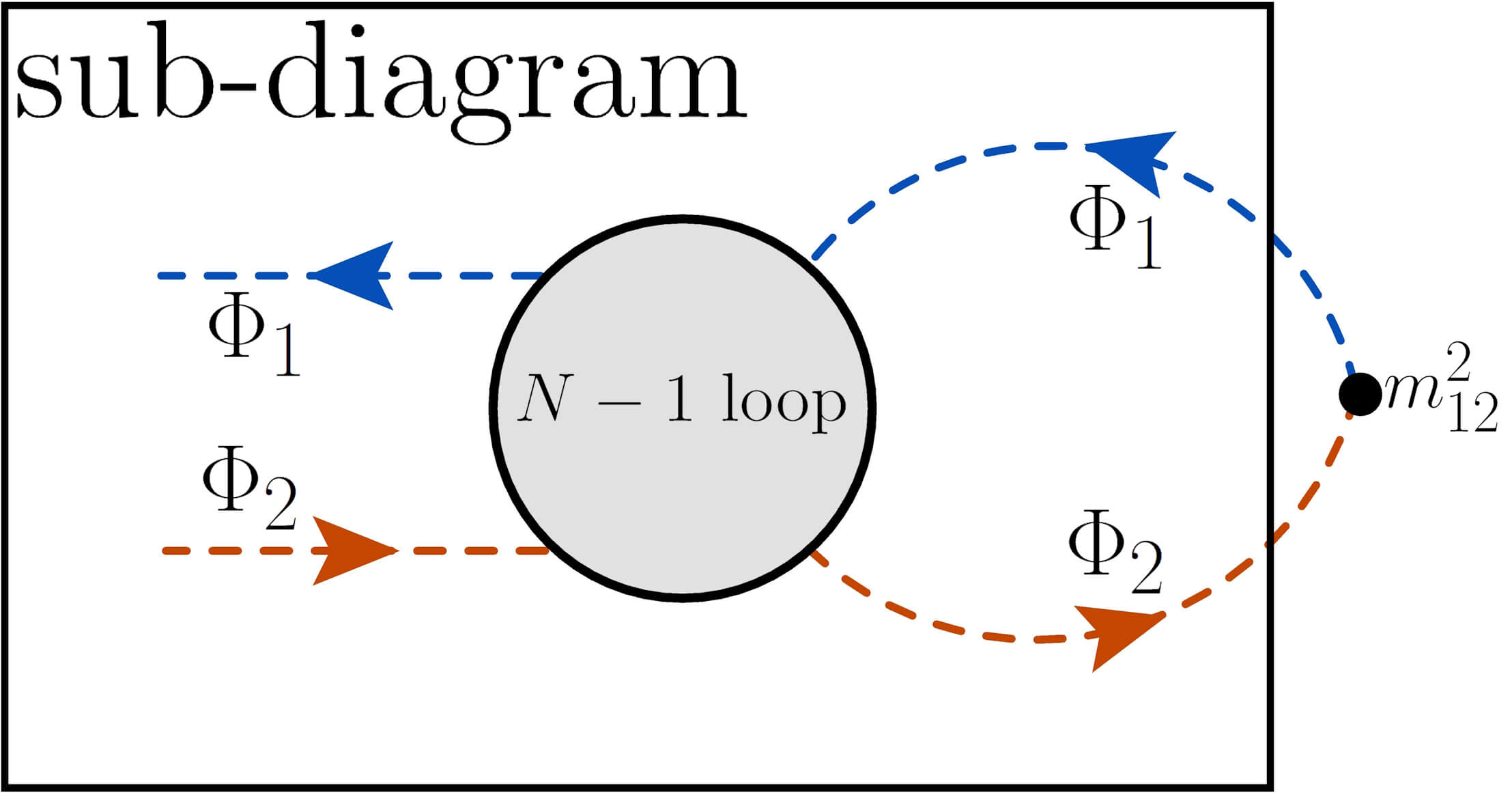}} \, \, \, \, \,  \resizebox{0.33\linewidth}{!}{ \includegraphics{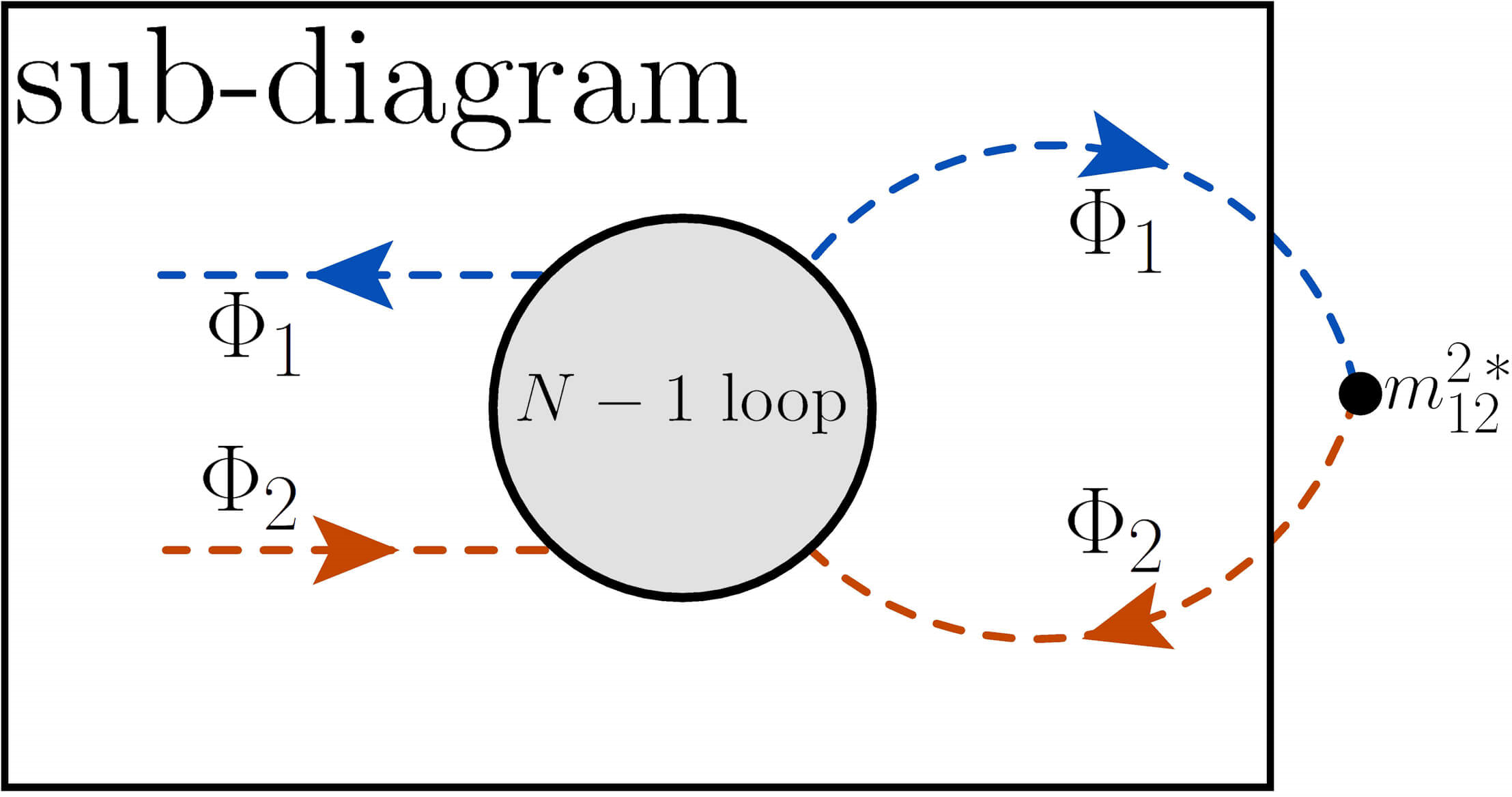}}
\caption{Subdiagram structure of all contributions to $\mathcal{O}_{12}$.  The diagram proportional to $m_{12}^2$ (left) always involves a Hermitian subdiagram. The diagram proportional to $m_{12}^{2*}$ (right) involves the non-Hermitian subdiagram corresponding to $\mathcal{O}_5$. }
\label{fig:m12sq-sub}
\end{figure}

We immediately observe that any $N$-loop diagram that contributes to $\mathcal{O}_{12} \equiv \Phi_1^{\dagger} \Phi_2$ must have the structure of an $(N-1)$-loop subdiagram with two legs connected to a $m_{12}^2$ or $m_{12}^{2*}$ mass insertion, as shown in Fig.~\ref{fig:m12sq-sub}.  The subdiagram connected to $m_{12}^2$ is Hermitian and corresponds to $\mathcal{O}_4 \equiv \Phi_1^{\dagger} \Phi_2 \Phi_2^{\dagger} \Phi_1$; we demonstrated in Sec.~\ref{sec:6loopTypeI} that such a four-scalar Hermitian subdiagram structure always leads to a cancellation of the imaginary divergent parts of the full $N$-loop diagram. We are then left with the non-Hermitian $(N-1)$-loop subdiagram connected to $m_{12}^{2*}$ shown in the right panel of Fig.~\ref{fig:m12sq-sub}.  Since this subdiagram corresponds to $\mathcal{O}_5$, it becomes clear that an imaginary divergent contribution to $m_{12}^2$ always appears at one higher loop order than the imaginary divergent contribution to $\lambda_5$ and is obtained simply by dressing the $\mathcal{O}_5$ operator with an $m_{12}^{2*}$ insertion.  

The fact that the imaginary divergent contribution to $m_{12}^2$ remains zero at the first loop order at which an imaginary divergent contribution to $\lambda_5$ appears ensures that the phase of $\lambda_5$ cannot be trivially rotated away at that order, and therefore that the CP violation that leaks into the 2HDM from the CKM matrix is indeed physical.

%%%%%%%%%%%%%%%%%%%%%%%%%%%%%%%%%%%%%%%%%%%
\section{Conclusion}
\label{sec:conclusion}
%%%%%%%%%%%%%%%%%%%%%%%%%%%%%%%%%%%%%%%%%%%

In this paper, we studied the conditions under which the CP violation in the CKM matrix can give rise to imaginary divergent radiative corrections to parameters in the scalar potential of the real 2HDM with natural flavor conservation.  Working in the unbroken phase, we demonstrated that such corrections cannot arise from diagrams involving Yukawa insertions alone because the Yukawa couplings preserve a global $U(1)_{PQ}$ symmetry under which the potentially-complex $\lambda_5$ and $m_{12}^2$ Lagrangian parameters are not invariant.  Breaking this symmetry requires an insertion of the tree-level coupling $\lambda_5$, pushing the calculation to the six-loop level.  This also implies that the 2HDM in which natural flavor conservation is enforced by a softly-broken $U(1)_{PQ}$ symmetry, which requires $\lambda_5 = 0$ at tree level, is guaranteed to be free from imaginary divergent radiative corrections to the scalar potential.  The importance of natural flavor conservation to these results is highlighted by the fact that in the general 2HDM without natural flavor conservation, the CP violation from the Yukawa couplings gives rise to imaginary divergent radiative corrections to the scalar potential parameters already at one loop.

We further showed that the six-loop diagrams that would be expected to generate imaginary divergent corrections instead possess enhanced symmetries in both the type I and II 2HDMs, which lead to complete cancellations of the imaginary divergent contributions at this order. Extending the analysis to seven loops, we identify the classes of diagrams that violate the enhanced symmetries that were present at six loops.  We find no evidence at seven loops for any remaining mechanism by which the imaginary divergent contributions would cancel.  The expected RG evolution for $\Im(\lambda_{5})$ starting at the seven-loop level can then be written as
\begin{align}
 	\frac{d \, {\rm Im}(\lambda_{5})}{d \ln \mu} = 
 	\frac{ \lambda_5 {\rm Im}(\mathcal{J})}{(16 \pi^2)^7} \begin{cases} 
      \left[ a^{\lambda} (\lambda_1 - \lambda_2) 
			+ a^{g^{\prime}} g^{\prime 2} 
			+ a^y (y_t^2 - y_b^2 + \ldots) \right]   &  
       \, \text{(type II)} \\
      \\
      \left[ b^{\lambda_3} \lambda_3 + b^{\lambda_4} \lambda_4 + b^{g^{\prime}} g^{\prime 2} + b^g g^2 \right]  &    \, \text{(type I).} \\
   \end{cases}
\end{align}
We also show that the imaginary divergent radiative corrections to $m_{12}^2$ are obtained starting at eight loops by dressing the seven-loop correction to $\lambda_5$ with an additional $m_{12}^{2*}$ insertion.

Interpreted in the context of RG evolution of the scalar potential parameters from an accidentally real potential at some high scale, these seven-loop CP-violating effects are so tiny as to be entirely phenomenologically irrelevant. Indeed, \emph{finite} imaginary contributions to the scalar potential, which we expect to appear starting at lower orders, will likely be numerically more important.
However, unless some as-yet-unidentified symmetry forces the imaginary divergent contributions to the scalar potential parameters to be zero to \emph{all} orders, the real 2HDM with softly-broken $Z_2$ symmetry is formally inconsistent under renormalization because it lacks counterterms for the imaginary parts of $\lambda_5$ and $m_{12}^2$.

%%%%%%%%%%%%%%%%%%%%%%%%%%%%%%%%%%%%%%%%%%%%
\acknowledgments%\section*{Acknowledgement}
C.H.L.\ thanks the organizers of the 2022 Workshop on Multi-Higgs Models in Lisbon for creating a stimulating environment in which the idea for this project arose. We also thank Pedro Ferreira, Duarte Fontes, Shinya Kanemura, David Morrissey, Rui Santos, and Daniel Stolarski for helpful conversations.
This work was supported by the Natural Sciences and Engineering Research Council of Canada (NSERC). C.H.L.\ is also supported by TRIUMF, which receives federal funding via a contribution agreement with the National Research Council (NRC) of Canada.
%%%%%%%%%%%%%%%%%%%%%%%%%%%%%%%%%%%%%%%%%%%%

\appendix

%%%%%%%%%%%%%%%%%%%%%%%%%%%%%%%%%%%%%%%%%%%
\section{Leaks of CP violation at one loop in the general 2HDM}
\label{sec:general2HDM}
%%%%%%%%%%%%%%%%%%%%%%%%%%%%%%%%%%%%%%%%%%%

\begin{figure}[h!]
 \resizebox{0.28\linewidth}{!}{ \includegraphics{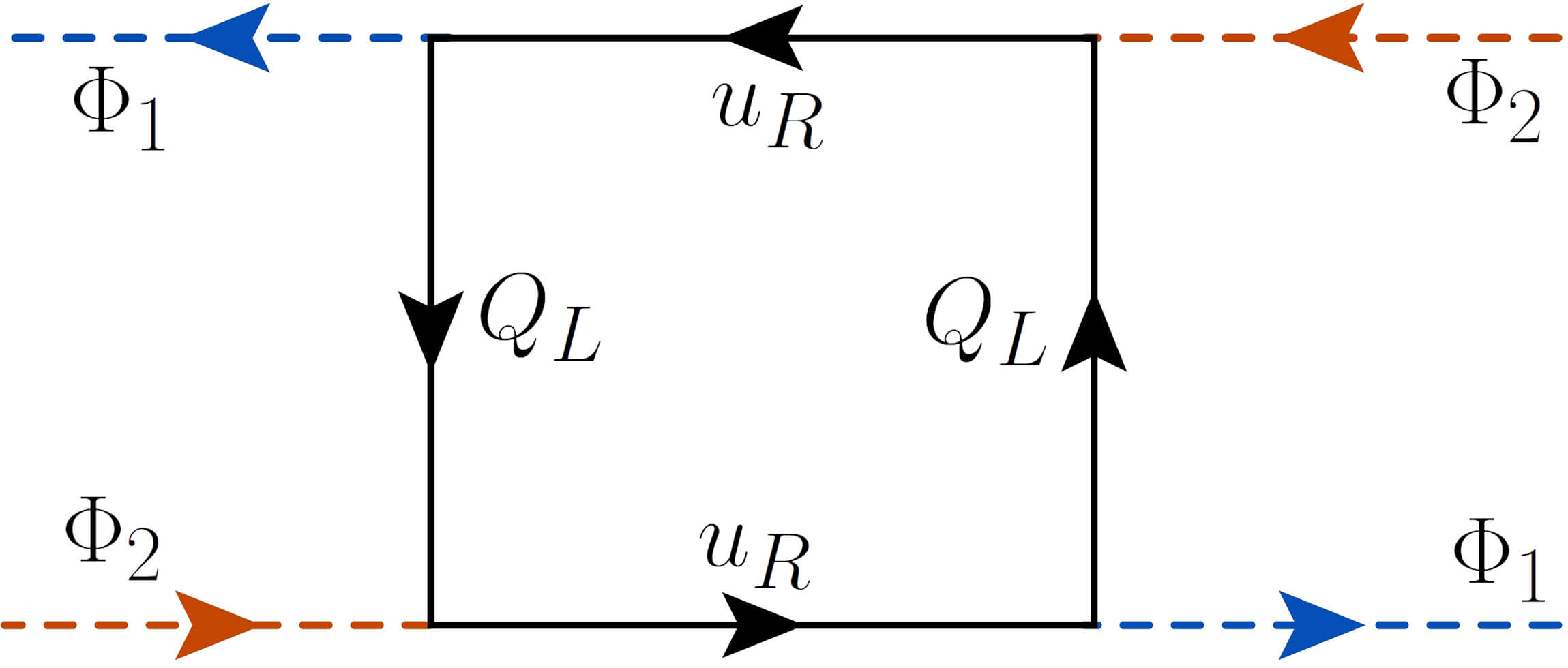}}\, \, \, \, \,
  \resizebox{0.28\linewidth}{!}{ \includegraphics{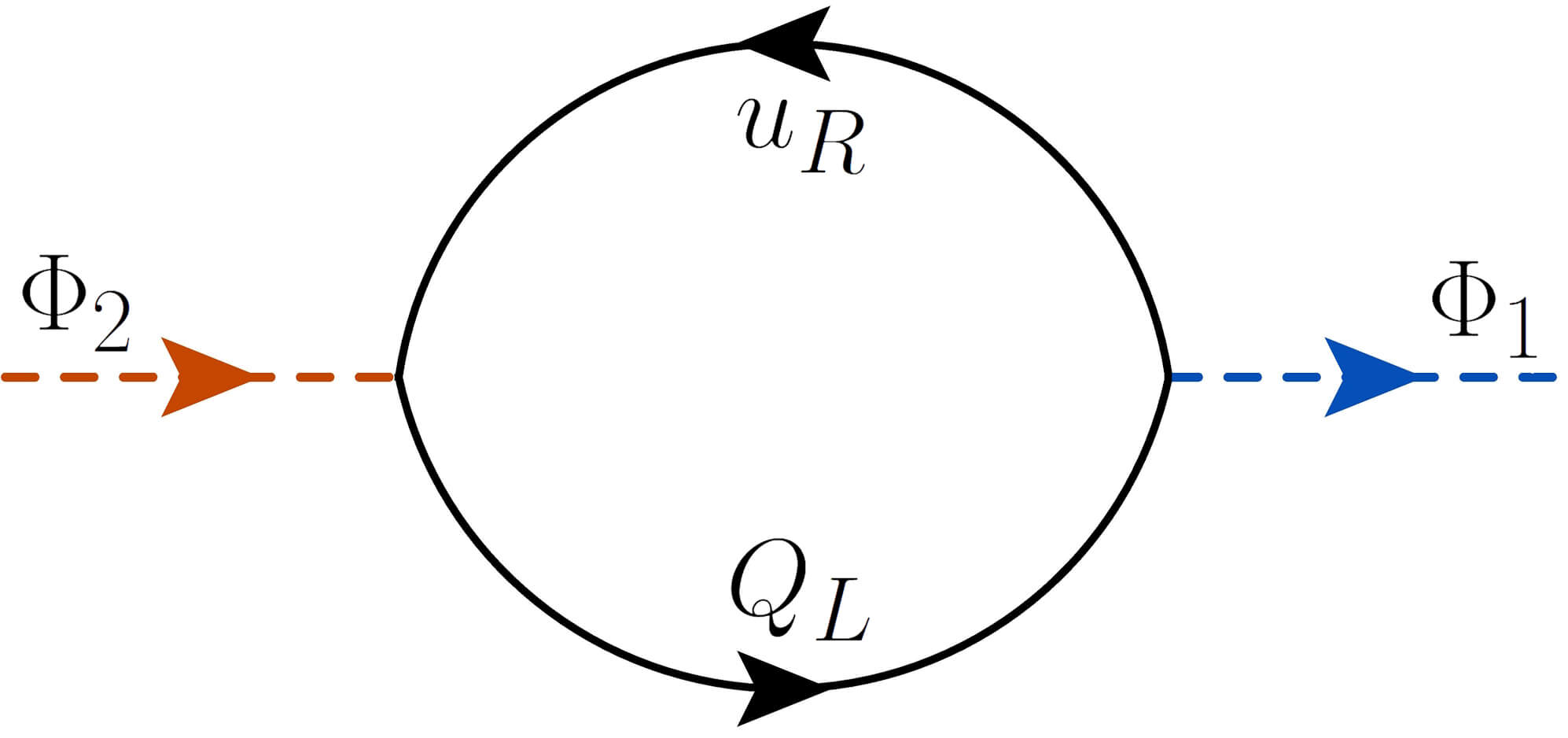}}
\caption{Sample diagrams contributing to $\lambda_5$ and $m_{12}^2$ at one-loop level in the general 2HDM.}
\label{fig:genYUK}
\end{figure}

An illustrative example highlighting the risk of assuming CP conservation in the scalar potential is provided by the general 2HDM of Sec.~\ref{sec:gen2HDM}, in which both Higgs doublets couple to all quarks. We work in the unbroken phase and consider the one-loop contributions to $\mathcal{O}_5 \equiv \Phi_1^{\dagger} \Phi_2 \Phi_1^{\dagger} \Phi_2$ and $\mathcal{O}_{12} \equiv \Phi_1^{\dagger} \Phi_2$. For simplicity, we consider only the diagrams involving up-type quarks shown in Fig.~\ref{fig:genYUK}. There is an analogous contribution from down-type quarks.

The diagrams of Fig.~\ref{fig:genYUK} involve the Yukawa couplings of $u_R$ to both $\Phi_1$ and $\Phi_2$. The UV-divergent parts of the two diagrams in Fig.~\ref{fig:genYUK} are easy to compute and depend on the Yukawa couplings according to 
\begin{equation}
	\left(\delta \lambda_{5}\right)_{\text{div}} \propto {\rm Tr}\left( Y_{u}^{(1)} Y_{u}^{(2) \, \dagger} Y_{u}^{(1)} Y_{u}^{(2) \, \dagger} \right), \qquad
	\left(\delta m_{12}^{2}\right)_{\text{div}} \propto {\rm Tr} \left( Y_{u}^{(1)} Y_{u}^{(2) \, \dagger} \right),
	\label{eq:gendiv}
\end{equation}
where we use the Yukawa coupling matrices as defined in Eq.~(\ref{eq:genYuk}).
Both of these traces have a nonzero imaginary part in general.

To complete the proof, we must demonstrate that the complex phases of $\lambda_5$ and $m_{12}^2$ induced at one loop cannot be rotated away by field definitions. A physically-meaningful phase exists only if ${\rm Im}[(m_{12}^{2*})^2 \lambda_5] \neq 0$.  From the structure of the Yukawa traces in Eq.~(\ref{eq:gendiv}), we immediately see that we can rotate away the phases if $Y_{u}^{(2)} = \zeta_u Y_{u}^{(1)}$, where $\zeta_u$ is a complex number. This relationship between $Y_{u}^{(2)}$ and $Y_{u}^{(1)}$ is the condition required for {\it Yukawa alignment}~\cite{Pich:2009sp}, which provides an alternative to natural flavor conservation for eliminating tree-level flavor-changing neutral Higgs couplings. However, it has been demonstrated that the Yukawa-alignment condition is violated by RG evolution of the Yukawa couplings~\cite{Ferreira:2010xe}, so it cannot be relied upon to eliminate the need for CP-violating counterterms in the scalar potential.

We thus demonstrate that divergent CP violation leaks into the scalar potential of the \emph{general} 2HDM from the Yukawa couplings of the quarks to both doublets already at the one-loop level. Therefore, the imposition of CP conservation on the scalar potential is inconsistent under renormalization. We can attribute the appearance of this divergent CP violation at one loop to the absence of a $U(1)_{PQ}$ symmetry in the Yukawa Lagrangian of the general 2HDM.

%%%%%%%%%%%%%%%%%%%%%%%%%%%%%%%%%%%%%%%%%%%%
\section{Vanishing of the $A^0$ tadpole at three loops}
\label{sec:tadpole}
%%%%%%%%%%%%%%%%%%%%%%%%%%%%%%%%%%%%%%%%%%%%
\begin{figure}[h!]
\resizebox{0.28\textwidth}{!}{\includegraphics{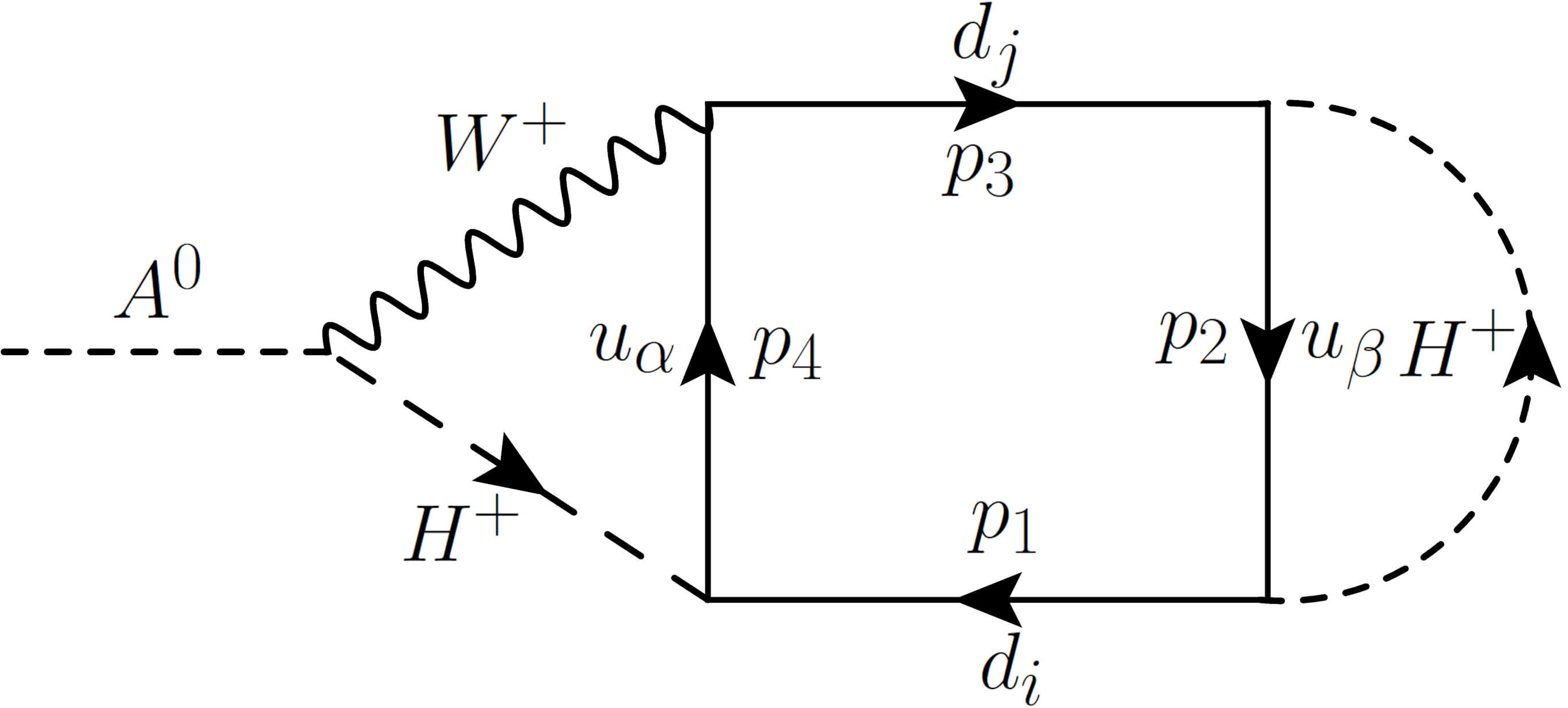}} \, \, \, \, \, \resizebox{0.28\textwidth}{!}{\includegraphics{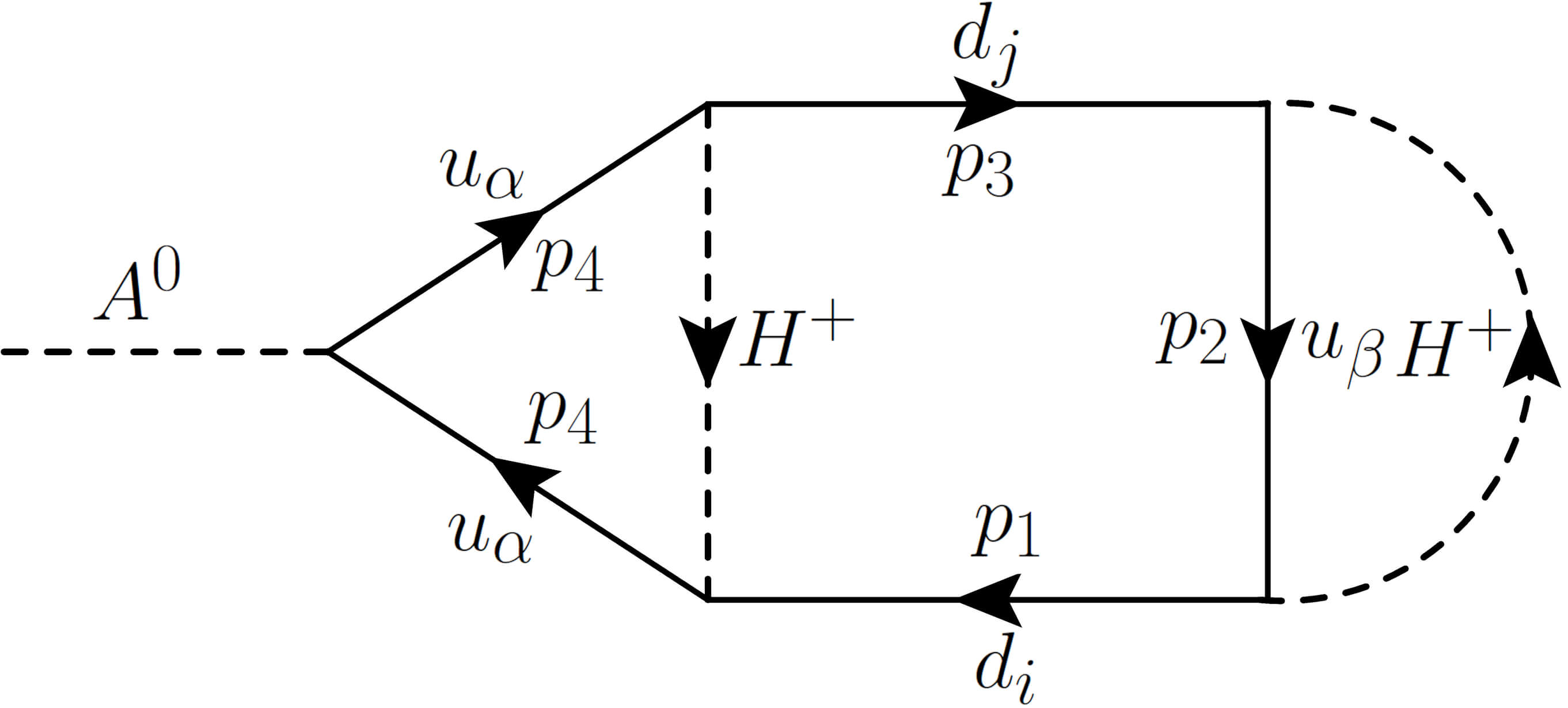}}
\caption{Two sample diagrams contributing to the $A^0$ tadpole at three loops.  Note that momentum conservation requires $p_1 = p_3$.}
\label{fig:tadpolep1p3}
\end{figure}

In this appendix, we confirm the result of Fontes et al.~\cite{Fontes:2021znm} that the leading $1/\varepsilon^3$ pole of the CP-violating $A^0$ tadpole is zero at three loops after summing over quark generations, and show that the subleading divergences and finite part of the tadpole are also zero at this order. We work entirely in the quark mass basis in the broken phase of the type II 2HDM; the corresponding results in the type I 2HDM differ only in the replacement of $\tan\beta$ with $-\cot\beta$ in the charged Higgs and $A^0$ couplings proportional to the down-type quark masses. 

\begin{figure}[h!]
\resizebox{0.28\textwidth}{!}{\includegraphics{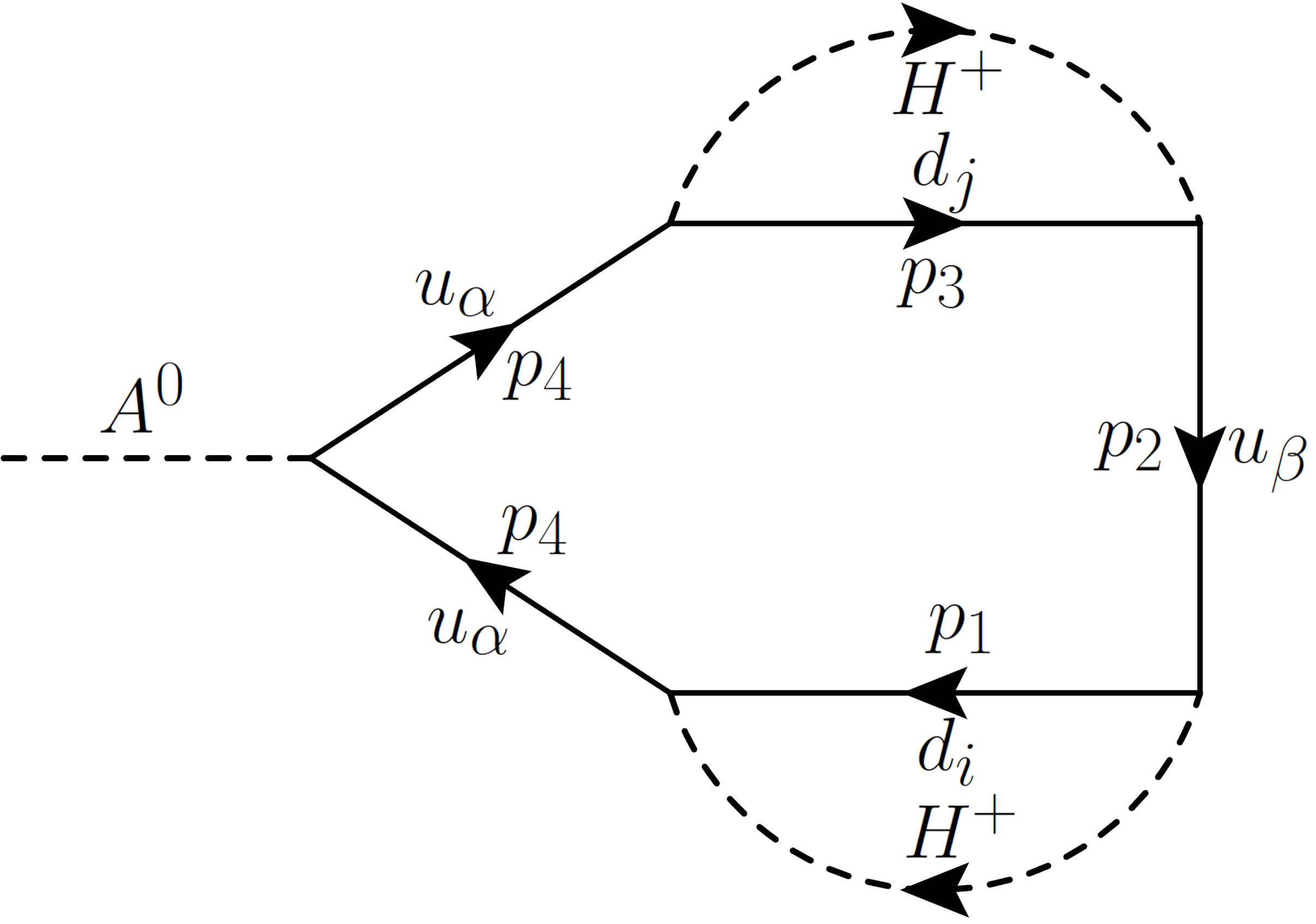}}
\caption{Another sample diagram contributing to the $A^0$ tadpole at three loops.  Note that momentum conservation requires $p_2 = p_4$.}
\label{fig:tadpolep2p4}
\end{figure}

The three-loop diagrams contributing to this process all involve a closed quark loop containing four charged boson ($H^{\pm}$, $G^{\pm}$, or $W^{\pm}$) vertices to yield four powers of the CKM matrix. The pseudoscalar $A^0$ can be attached directly to one of the quark lines or to a pair of charged bosons. Some sample diagrams are shown in Figs.~\ref{fig:tadpolep1p3} and \ref{fig:tadpolep2p4}, in which we have denoted the up-type quark mass eigenstates by $\alpha$ and $\beta$ and those of the down-type quarks by $i$ and $j$.
We focus on the diagrams with fermion flow and electric charge flow as shown in Figs.~\ref{fig:tadpolep1p3} and \ref{fig:tadpolep2p4}. Reversing the fermion flow gives minus the Hermitian conjugate of the original diagrams (due to the odd parity of the $A^0$ coupling) and thereby selects out the imaginary part of the trace over CKM matrix elements, as expected. A second set of diagrams is obtained by interchanging $u$ and $d$; their calculation follows the same method.

We proceed to write down the amplitude for each diagram, multiply out the couplings and the numerators of the propagators,\footnote{We work in dimensional regularization in the scheme in which $\gamma^5$ anticommutes with all of the other four gamma matrices.} and examine the sum over quark flavors before performing any loop momentum integrals. For each diagram, we find that the part of the amplitude that depends on the quark masses always takes the generic form
\begin{equation}
	\sum_{\alpha, \beta, i, j} V^*_{\alpha j} m^{n_1}_{\alpha} V_{\alpha i} m^{m_1}_i V^*_{\beta i} m^{n_2}_{\beta} V_{\beta j} m^{m_2}_j 
	\, \left[(p_1^2 - m_i^2)(p_2^2 - m_{\beta}^2)(p_3^2 - m_j^2)(p_4^2 - m_{\alpha}^2)\right]^{-1},
	\label{eq:tadmaster}
\end{equation}
where $V$ is the CKM matrix and each of the integers $n_1$, $n_2$, $m_1$, and $m_2$ can be equal to 0 or 2.

A couple of explanations are in order. First, momentum conservation requires the incoming four-momentum of $A^0$ to be zero. We have already taken this into account in the momentum labels.
Second, some of the diagrams contain two $u_{\alpha}$ propagators, yet only one propagator denominator involving $m_{\alpha}$ appears in Eq.~(\ref{eq:tadmaster}). This is because the algebra around the $A^0$ vertex takes the form
\begin{equation}
	\frac{\ps_4 + m_{\alpha}}{p_4^2 - m_{\alpha}^2} y_{\alpha} \gamma^5 \frac{\ps_4 + m_{\alpha}}{p_4^2 - m_{\alpha}^2}
	= - \frac{(p_4^2 - m_{\alpha}^2)}{(p_4^2 - m_{\alpha}^2)^2} y_{\alpha} \gamma^5,
\end{equation}
where $y_{\alpha} \propto m_{\alpha}$ is the $A^0$ coupling strength, and thus one of the propagator denominators cancels.

The sum over quark generations in Eq.~(\ref{eq:tadmaster}) is complicated by the presence of the propagator denominators. As a first step, consider what would happen if the masses in each propagator denominator were dropped, i.e., if we were to take $p_i^2 \gg m^2$. This corresponds to taking the most divergent piece of the loop integral. In this case, the sum over quark generations yields a purely real result. We can see this as follows.  
First, consider that one of the quark masses does not appear in the sum; for concreteness, we pick $m_i$. Then the sum over $i$ gives $\sum_i V_{\alpha i} V^*_{\beta i} = \delta_{\alpha \beta}$ because the CKM matrix is unitary; using the Kronecker delta to perform the sum over $\alpha$ then leaves $V^*_{\beta j} V_{\beta j}$ times some powers of masses, which is purely real.
Second, consider the case in which $m_i$ and $m_j$ appear with the same power. Then, the dummy indices $i$ and $j$ can be interchanged, which interchanges $V \leftrightarrow V^*$ in the sum, and the result is again purely real. The same conclusion holds if $m_{\alpha}$ and $m_{\beta}$ appear with the same power. Since each of the exponents $n_1$, $n_2$, $m_1$, and $m_2$ is equal to either 0 or 2 in any contributing diagram, the mass dependence of the denominators of the propagators \emph{must} play a role if the imaginary part is not to cancel immediately.

To pull powers of quark masses from the propagator denominators into the numerator, we perform an expansion in powers of $m^2/p^2$,
\begin{equation}
	\frac{1}{p^2 - m^2} = \frac{1}{p^2} \left[ 1 + \frac{m^2}{p^2} + \frac{m^4}{p^4} + \cdots \right].
\end{equation}
Consider first the diagrams of the type shown in Fig.~\ref{fig:tadpolep1p3}, in which $p_1 = p_3$. The numerators of these diagrams contain terms, each of which is proportional to $m_i^0 m_j^0$, $m_i^2 m_j^0$, $m_i^0 m_j^2$, or $m_i^2 m_j^2$. To obtain an imaginary contribution from the sum over CKM matrices, we need, at a minimum, $m_i^2 m_j^4$ or $m_i^4 m_j^2$. Let us aim for $m_i^2 m_j^4$. Expanding the $d_i$ and $d_j$ propagators as described and starting from, e.g., $m_i^{n_1} m_j^{n_2}$ in the numerator, the relevant piece is
\begin{equation}
	m_i^{n_1} m_j^{n_2} \, \frac{1}{p_1^2 p_3^2} \left( \frac{m_i^2}{p_1^2} \right)^{2-n_1}
		\left( \frac{m_j^2}{p_3^2} \right)^{4-n_2}
		= m_i^2 m_j^4 \, \frac{1}{p_1^2 p_3^2} \frac{1}{(p_1^2)^{2-n_1} (p_3^2)^{4-n_2}}.
\end{equation}
But for each such term in the expansion of the propagators, there is another term multiplying the same overall coefficient,
\begin{equation}
	m_i^{n_1} m_j^{n_2} \, \frac{1}{p_1^2 p_3^2} \left( \frac{m_i^2}{p_1^2} \right)^{4-n_1}
		\left( \frac{m_j^2}{p_3^2} \right)^{2-n_2}
		= m_i^4 m_j^2 \, \frac{1}{p_1^2 p_3^2} \frac{1}{(p_1^2)^{4-n_1} (p_3^2)^{2-n_2}}.
\end{equation}
Since we are considering the diagrams in which $p_1 = p_3$, the momentum-dependent parts of these two terms are identical. They differ only by exchange of the dummy indices $i \leftrightarrow j$, which interchanges $V \leftrightarrow V^*$ in the sum over quark generations. Thus, the imaginary part of the product of CKM matrices cancels.
An identical argument can be constructed for the diagrams of the type shown in Fig.~\ref{fig:tadpolep2p4}, in which $p_2 = p_4$, by considering the powers of $m_{\alpha}$ and $m_{\beta}$ in the numerator.  

The pairwise cancellation of the imaginary part of the sum over CKM matrices between pairs of terms in the expansion of the propagators happens to all orders in the expansion. This demonstrates that all contributions to the $A^0$ tadpole (including the finite parts) cancel in the sum over quark masses diagram-by-diagram and before any loop integrals are performed, due to the structure of the trace over generation indices.

The cancellation demonstrated here could potentially be circumvented by going to a higher loop order (so that the number of quark mass insertions in the numerator is sufficient to avoid expanding propagators) or by considering a more complicated momentum flow, such as in $A^0$--$H^0$ mixing (so that the quark propagators all carry different momenta). Rather than pursuing these avenues, we found it more straightforward in the main body of this paper to work in the unbroken phase of the 2HDM and target the imaginary divergent contribution to the scalar potential parameter $\lambda_5$.

%%%%%%%%%%%%%%%%%%%%%%%%%%%%%%%%%%%%%%%%%%%%

\bibliography{bib2HCPV}

\end{document}